\documentclass{article}

\usepackage{PRIMEarxiv}

\usepackage[utf8]{inputenc} 
\usepackage[T1]{fontenc}    
\usepackage{hyperref}       
\usepackage{url}            
\usepackage{booktabs}       
\usepackage{nicefrac}       

\usepackage{lipsum}
\usepackage{fancyhdr}       
\usepackage{graphicx}       
\graphicspath{{media/}}     
\usepackage{amssymb}
\usepackage{amsfonts, amsthm, amsmath}
\usepackage{algorithm, setspace}
\usepackage{algpseudocode}
\usepackage[usenames,dvipsnames]{xcolor}
\usepackage{color}

\title{BCM-Broadcast: A Byzantine-Tolerant Causal Broadcast Algorithm for Distributed Mobile Systems
\thanks{\textit{\underline{Citation}}: 
\textbf{Authors. Title. Pages.... DOI:000000/11111.}} 
}

\author{
  Leila NamvariTazehkand \\
  Department of Computer Engineering\\
  Faculty of Electrical and Computer Engineering \\
  University of Tabriz \\
  Tabriz, East Azerbaijan\\
  Iran\\
\texttt{l.namvari@tabrizu.ac.ir} \\
   \And
  Saied Pashazadeh\\
  Department of Information Technology\\
  Faculty of Electrical and Computer Engineering \\
  University of Tabriz \\
  Tabriz, East Azerbaijan\\
  Iran\\
  Corresponding author:\\ \texttt{pashazadeh@tabrizu.ac.ir} \\
   \And
  Ali Ebnenasir \\
  Department of Computer Science \\
  Michigan Technological University \\
   Houghton, MI, 49931\\
    USA\\
  \texttt{aebnenas@mtu.edu} \\
}

\begin{document}
\maketitle

\begin{abstract}
    
This paper presents an algorithm, called {\it BCM-Broadcast}, for the implementation of causal broadcast in distributed mobile systems in the presence of Byzantine failures. The {\it BCM-Broadcast} algorithm simultaneously focuses on three critical challenges in distributed systems: Byzantine failures, Causality, and Mobility. We first present a hierarchical architecture for {\it BCM-Broadcast}. Then, we define twelve properties for {\it BCM-Broadcast}, including validity, integrity, termination, and causality. We then show that {\it BCM-Broadcast} satisfies all these properties. We also prove the safety of {\it BCM-Broadcast}; i.e., no healthy process delivers a message from any Byzantine process, assuming that the number of Byzantine processes is less than a third of the total number of mobile nodes. Subsequently, we show that the message complexity of {\it BCM-Broadcast} is linear. Finally, using the Poisson process, we analyze the probability of the violation of the safety condition under different mobility scenarios. 
\end{abstract}

\newtheorem{Lma}{Lemma}[section]
\newtheorem{Theo}[Lma]{Theorem}
\section{Introduction}
\label{sec:Intro}

This paper presents a novel algorithm, called BCM-Broadcast, for ensuring the causal order of messages in distributed mobile systems where  nodes may be prone to Byzantine failure. With the advent of distributed mobile systems (e.g., connected vehicles), the presence of Byzantine nodes is inevitable; thus, having a causal broadcast algorithm is especially important for such systems. Due to node mobility and Byzantine failures, designing such an algorithm is a  challenging tasks, let alone proving its correctness. To address the design problem, we first present a hierarchical architecture and then a layered algorithm called \textit{BCM-Broadcast} (Byzantine, Causality, and Mobility-Broadcast), which simultaneously manages all three mentioned problems. For the proof of correctness of the proposed algorithm, we first specify the properties of the algorithm such as validity, integrity, termination, causality, and safety. Then, we show that BCM-Broadcast satisfies all its required properties.

Most existing algorithms for the causal delivery of messages have focused on ensuring either (i) the causal order in distributed mobile systems, or (ii) the causal order of messages in the presence of Byzantine processes, but not both. For example, Benzaid \textit{et al.} \cite{30} present a protocol called \textit{Mobi-causal} for implementing causal ordering in mobile computing systems, which is suitable for unicast communication. In a separate research \cite{31}, they present another protocol called \textit{Bmobi-causal} inspired by \cite{30} to ensure the causal ordering of messages for broadcast communications. N´edelec \textit{et al.} \cite{19} focus on the scalability of causal broadcast algorithms in dynamic and large systems. Guidec \textit{et al.} \cite{21}, guarantee the causal delivery of messages in opportunistic networks, wherein one of their algorithms, they consider a lifetime for each broadcast message. Ohori \textit{et al.} \cite{1} introduce a distributed mobile system that contains \textit{Mobile Hosts} (MH) and \textit{Mobile Support Station} (MSS), where the MSS nodes are stationary and the MH nodes are mobile. They also propose an efficient causal broadcast protocol for this system. Auvolat \textit{et al.} \cite{2} present a causal broadcast algorithm in the presence of Byzantine processes, where the participating processes are non-mobile. Misra \textit{et al.} \cite{14} investigate the protocol presented in \cite{11} to investigate the Byzantine tolerance of this algorithm. None of the causal broadcast algorithms presented for mobile systems have considered the participating nodes to be Byzantine, and none of the causal broadcast algorithms presented in the presence of Byzantine processes have considered the participating nodes to be mobile.

In this paper, we first present a new architecture inspired by \cite{1}, which contains MH and MSS nodes. We consider each MSS node as a stationary and always non-faulty process. In contrast to an MSS node, an MH node can move and is prone to Byzantine failure. Each MSS node communicates with a set of MH nodes in its radio range as a {\it group}, and each MH node directly exchanges messages only with the MH nodes and the MSS node of its group. The presented architecture supports hierarchical communication; thus, the message exchange occurs among an MSS and its group's MH nodes in the lower layer. Inter-group message transmission occurs from the MSS node of one group to the MSS node of another group in a higher layer. The number of MH and MSS nodes in the system is fixed. The total number of MH nodes that can be Byzantine in the entire system is less than one-third of the total number of MH nodes. Moreover, the total number of MH nodes that can be Byzantine in a group must be less than one-third of the total number of MH nodes present in that group.

The proposed BCM-Broadcast algorithm includes the following layers: the \textit{Byzantine failure management layer}, the \textit{causal delivery management layer}, and the \textit{mobility management layer}, which respectively manage the Byzantine failure of the MH nodes, the causal delivery of messages, and the handoff of MH nodes from one group to another. We then prove that the BCM-Broadcast algorithm satisfies its correctness properties related to validity, integrity, causality, and termination. In addition, we investigate the types of events that increase the number of Byzantine MH nodes in a group and then analyze the probability of violation of the correctness conditions of the algorithm using the Poisson process. Finally, we calculate the number of communication steps and the number of messages exchanged until an arbitrary node delivers a message and show that the message complexity of our algorithm is linear. To the best of our knowledge, the BCM-Broadcast algorithm is the first one that simultaneously manages the three important challenges of Byzantine failure, causal delivery, and mobility in distributed mobile systems.

\textbf{Organization of the paper}. Section \ref{sec:basics} provides preliminary knowledge about the mobile distributed systems architecture and basic concepts.  Section \ref{sec:def} defines the properties of the BCM-Broadcast algorithm.
Section \ref{sec:alg} presents the BCM-Broadcast algorithm and its layers. Section \ref{sec:corecct} proves the properties of the BCM-Broadcast algorithm. Section \ref{sec:analysis} analyzes the probability of the violation of the safety condition. Section \ref{sec:related} discusses related works on the causal broadcast algorithm for distributed mobile systems and Byzantine-tolerant causal broadcast algorithms. Finally, Section \ref{sec:concl} makes concluding remarks and discusses future work.
\section{Basic Concepts}\label{sec:basics}
In this section, we first represent the architecture of the distributed mobile system presented in \cite{1} and then present a new architecture inspired by \cite{1} where the mobile nodes are prone to Byzantine failure. In addition, we improve the definitions in \cite{2} to capture the mobility of MH nodes. The algorithm presented in this paper is inspired by their approach to managing the handoff of processes \cite{1} and guaranteeing causal order in the presence of Byzantine failures \cite{2}. Section \ref{sec2.1} presents a novel architecture for mobile distributed systems in the presence of Byzantine processes. Section \ref{sec2.2} presents basic concepts. 


\subsection{Distributed Mobile System}\label{sec2.1}
For the presentation of new architecture, we basically follow \cite{1}. Based on \cite{1}, a distributed mobile system $S$ consists of two types of components: MSS and MH, where MSS represents a collection of processes referred to as mobile support stations (MSS nodes). MH encompasses a collection of processes known as mobile hosts (MH nodes). Let $N_{mss}$ and $N_{mh}$ denote the number of MSS and MH nodes, respectively. The number of MSS and MH nodes is limited and fixed. Let MSS represent $\left\{s_{1}, ..., s_{N_{mss}}\right\}$ and MH denote $\left\{h_{1}, ..., h_{N_{mh}}\right\}$. An MH node can leave a station and connect to another station. In \cite{1}, all MH and MSS nodes are assumed to be non-faulty processes.

\begin{figure}
\begin{minipage}[c]{0.5\linewidth}
\includegraphics[width=\linewidth]{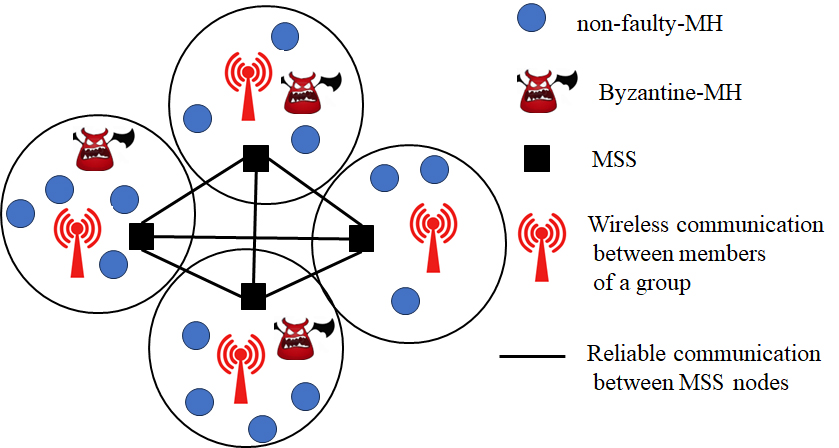}
\caption{A distributed mobile system}
\label{fig1}
\end{minipage}
\hfill
\begin{minipage}[c]{0.5\linewidth}
\includegraphics[width=\linewidth]{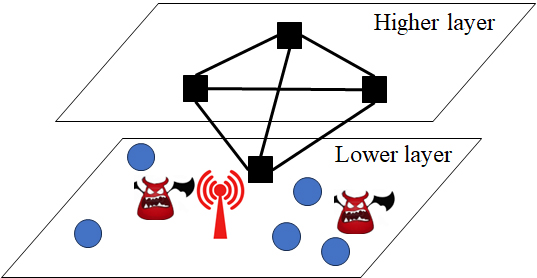}
\caption{The hierarchical architecture}
\label{fig2}
\end{minipage}%
\end{figure}

As shown in Figure \ref{fig1}, each MSS node communicates as a station with a set of MH nodes in its radio frequency (\textit{RF}) range, considered as its \textit{group}. Each MH node can directly exchange messages only with the MSS node and MH nodes in its group. All MSS nodes form a group called \textit{the MSS group}. Each MSS node broadcasts the exchanged messages in the MSS group with other MSS nodes, and they in turn can relay those messages to their own group. We assume that the communication between MSS nodes is performed through reliable bidirectional FIFO channels. The communication of an MH node with its group members is wireless and guarantees FIFO.

Figure \ref{fig2} illustrates a hierarchical communication in the desired system, which includes two layers called the lower and higher layers. Message exchange between the MSS nodes is performed at the higher layer, while message exchange between an MSS node and MH nodes in its group is performed at the lower layer. An MH node may be non-faulty or Byzantine, but an MSS node is always non-faulty. Each MH (respectively, MSS) node has an identifier, denoted $h_i$ (respectively,  $s_i$). Let $N$ denotes the total number of nodes (MSS and MH), where $N = N_{mss} + N_{mh}$.

\newtheorem{Def}{Definition}[section]
\begin{Def}[Handoff]\label{def3.1}
    \normalfont The procedure of transferring an MH from one group to another group is called \textit{Handoff}. In this procedure, an MH $h_i$ leaves an MSS $s_j$ and joins another MSS $s_k$. Figure \ref{fig4} illustrates the handoff steps for $h_i$. The MH $h_i$ goes through six different steps to handoff from $s_j$ to $s_k$, which include: 1) $h_i$ sends the messages \textit{disconnect} to $s_j$, 2) $h_i$ leaves $s_j$ and enters the RF range of $s_k$, 3) $h_i$ sends the messages \textit{requestMsg} to $s_k$, 4) $s_j$ sends information about $h_i$ to $s_k$ by sending the message \textit{removed}, 5) $s_k$ sends the message \textit{accept} to $s_j$, and 6) $s_k$ forwards the messages that $h_i$ has not yet delivered to $h_i$. 
\end{Def}


\subsection{Basic Definitions}\label{sec2.2}
In this section, we first state the concepts and definitions used throughout the paper. 
For the definition of new concepts, we basically follow \cite{2}. Auvolat \textit{et al.} \cite{2}, in one of their recent research, have presented, for the first time, an algorithm to ensure the causal ordering of messages in the presence of Byzantine processes.

\begin{Def}[Byzantine Failure]\label{Def2.1}
\normalfont A Byzantine process may crash, or may stop broadcasting a message and not deliver the message successfully, may refuse to deliver it when a message arrives or may refuse to send a confirmation message after a message arrives, and may broadcast duplicate or arbitrary messages in the system. A Byzantine process cannot manipulate the content of the message or the information in the message, and no Byzantine process can impersonate another process \cite{2}.
\end{Def}

Auvolat \textit{et al.} \cite{2} consider a system containing $n$ processes with up to $t$ Byzantine processes, where $t<n/3$. Their desired system is a static system where all processes are stationary. 
In our work, $t$ represents the total number of all Byzantine processes, where up to $t < \frac{N_{mh}}{3}$ of MH nodes can be Byzantine. We make the following assumptions about Byzantine processes:

\begin{itemize}
    \item Byzantine processes are scattered in different groups.
    \item The number of Byzantine processes in a group $j$ is less than a third of the total number of MH nodes in that group, which means $t_j < \frac{NMH_j}{3}$, where $NMH_j$ and $t_j$ respectively denote the number of MH nodes connected to MSS $s_j$ in the group and the number of Byzantine processes in that group, respectively. We call $t_j < \frac{NMH_j}{3}$ the \textit{t-condition}. 
\end{itemize}

Auvolat \textit{et al.} \cite{2} define the Byzantine Reliable Broadcast (BR-Broadcast) as a communication abstraction containing two operations: \textit{br-broadcast()} and \textit{br-deliver()}. It guarantees that non-faulty processes deliver the same set of messages despite the presence of Byzantine processes. Also, they introduce the Byzantine Causal Order Broadcast (BCO-Broadcast) with operations: \textit{bco-broadcast()} and \textit{bco-deliver()}. This algorithm is implemented in a layer on BR-Broadcast for causal broadcast in non-mobile networks in the presence of Byzantine processes. Both BR-Broadcast and BCO-Broadcast concepts are defined for stationary processes. Since we present our algorithm in a distributed system with a combination of stationary and mobile nodes, it is necessary to rewrite the BR-Broadcast algorithm according to the desired architecture and assumptions. 

\begin{Def}[BR-Broadcast for mobile systems]\label{def3.2}
    \normalfont The BR-Broadcast abstraction is a communication abstraction defined by two operations \textit{br-broadcast()} and \textit{br-deliver()}. These operations belong to the Byzantine failure management layer. In contrast to an MSS, an MH is subject to Byzantine failure. Thus, an MH node (non-faulty or Byzantine) can send a message by \textit{br-broadcast()}; also, an MSS and MH node can receive a message by \textit{br-deliver()} in the Byzantine failure management layer. The MH nodes deliver the messages that are br-broadcast in the Byzantine failure management layer through the MSS node after passing through the causal delivery management layer. The BR-Broadcast satisfies the following properties:
\end{Def}
\begin{enumerate}
    \item \textit{BR-Validity}: If an MSS node br-delivers a message $m$ from a non-faulty MH $h_i$, then $h_i$ must have br-broadcast $m$.
	\item \textit{BR-Integrity}: An MSS node br-delivers a message $m$ at most once from an MH $h_i$, even if $h_i$ has br-broadcast it more than once.
\end{enumerate}
\begin{Def}[C-Broadcast] \label{def3.3}
    \normalfont  The C-Broadcast abstraction is a communication abstraction based on the causal order introduced by two operations, \textit{c-broadcast()} and \textit{c-deliver()}. Thus, an MSS node sends a message to other MSS nodes with c-broadcast and delivers a message from another MSS with c-deliver in the causal delivery management layer. The C-Broadcast satisfies the following properties:
\end{Def}
\begin{enumerate}
     \item \textit{C-Validity}: If an MSS node c-delivers a message $m$ from an MSS node $s_j$, then $s_j$ must have c-broadcast $m$.
    \item \textit{C-Integrity}: An MSS node c-delivers a message $m$ at most once from $s_j$.
	\item \textit{C-Termination}: If an MSS node broadcast $m$ then all MSS nodes will eventually c-deliver $m$.
	\item \textit{C-Causality}: If an MSS node first c-delivers or c-broadcasts a message $m_1$ and then c-broadcasts the message $m_2$, no MSS node c-delivers $m_2$ before $m_1$.
\end{enumerate}

\section{Defining the Properties of BCM-Broadcast}\label{sec:def}
This section presents a layered algorithm called \textit{BCM-Broadcast} (Byzantine, Causality, and Mobility-Broadcast) that enables causal reliable broadcast in a network of mobile nodes where some might be Byzantine. The layer that manages the handoff of MH nodes is called the \textit{mobility management layer}. The layer that manages the Byzantine failure of the MH nodes is called the \textit{Byzantine failure management layer}, and the layer that guarantees the causal delivery of messages is called the \textit{causal delivery management layer}. Figure \ref{fig3} illustrates the relationship between these three layers. The causal delivery management layer is executed on the Byzantine failure management layer. However, the mobility management layer is executed independently of the other two layers but simultaneously with them.    We make the following four assumptions in the context of a distributed mobile system that is subject to Byzantine failures:
\begin{figure}
\begin{minipage}[c]{0.45\linewidth}
\includegraphics[width=\linewidth]{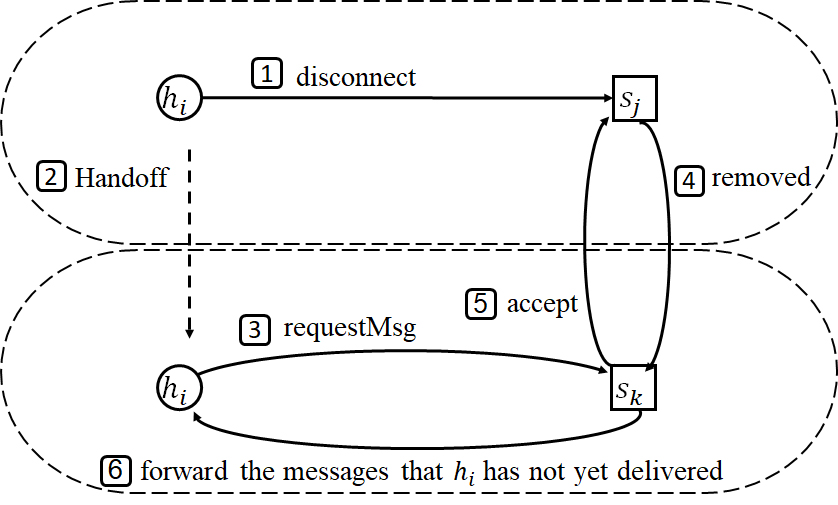}
\caption{The steps of handoff of $h_i$ from $s_j$ to $s_k$}
\label{fig4}
\end{minipage}%
\hfill
\begin{minipage}[c]{0.45\linewidth}
\includegraphics[width=\linewidth]{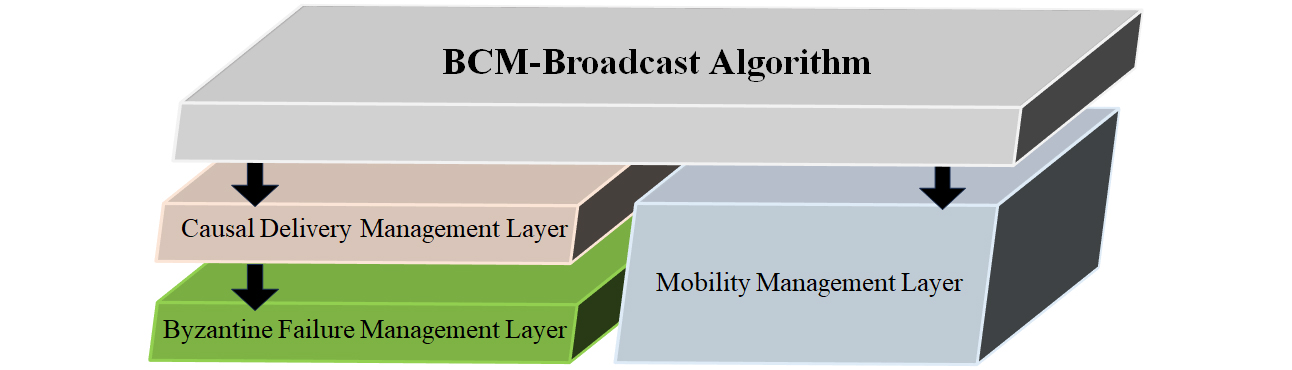}
\caption{The layers of the BCM-Broadcast algorithm}
\label{fig3}
\end{minipage}
\end{figure}

\begin{itemize}
    \item 	$h_i$ cannot connect to two MSS nodes simultaneously.
    \item   $h_i$ in transit (when it is not yet attached to a specific MSS) cannot broadcast or deliver a message.
    \item   A Byzantine process in the source group is also Byzantine in the destination group.
    \item   A non-faulty process in the source group is also non-faulty in the destination group.
\end{itemize}

\begin{Def}[BCM-Broadcast]\label{def3.4}
    \normalfont The BCM-Broadcast abstraction is a communication abstraction introduced by four operations, \textit{bcm-Hbroadcast()}, \textit{bcm-Hdeliver()}, \textit{bcm-Sbroadcast()} and \textit{bcm-Sdeliver()}. The operations of broadcasting and delivering a message by an MH $h_i$ are called \textit{bcm-Hbroadcast()} and \textit{bcm-Hdeliver()}, respectively. Also, an MSS $s_j$ performs broadcasting and delivering operations by \textit{bcm-Sbroadcast()} and \textit{bcm-Sdeliver()}.

Throughout the paper, the operations br-broadcast() and br-deliver() represent the primitive actions of broadcasting and delivering a message in the Byzantine failure management layer, respectively. Also, the operations c-broadcast() and c-deliver() represent the primitive actions of broadcasting and delivering a message in the causal delivery management layer, respectively. In addition, the operations
bcm-Hbroadcast() and bcm-Hdeliver() denote the primitive actions of broadcasting and delivering a message by an MH node in the application layer, as well as the operations bcm-Sbroadcast() and bcm-Sdeliver() representing the primitive actions of broadcasting and delivering a message by an MSS node in the application layer. The BCM-Broadcast is introduced with the following properties:
\end{Def}
\begin{enumerate}
    \item 	\textit{BCM-Validity 1}: If an MSS node bcm-Sdelivers a message $m$ from an MSS node $s_j$, then $s_j$ has bcm-broadcast $m$.
	\item  \textit{BCM-Validity 2}: If a non-faulty MH node $h_i$ bcm-Hdelivers a message $m$ from an MSS node $s_j$, then $s_j$ forwarded $m$ to $h_i$.
	\item  \textit{BCM-Validity 3}: If a non-faulty MH node (respectively, MSS node) bcm-Hdelivers (respectively, bcm-Sdelivers) a message $m$ from an MH node $h_i$, then $h_i$ bcm-Hbroadcast $m$.
	\item  \textit{BCM-Integrity 1}: A non-faulty MH node (respectively, MSS node) bcm-Hdelivers (respectively, bcm-Sdelivers) a message $m$ at most once.
	\item  \textit{BCM-Integrity 2}: If a message $m$ is bcm-Sbroadcast by MSS $s_j$ before $h_i$ is disconnected from a group and the same message is bcm-Sbroadcast by MSS $s_k$ after $h_i$ is connected to the group of $s_k$, then $h_i$ bcm-Hdelivers message $m$ at most once.
	\item  \textit{BCM-Termination 1}: If a non-faulty MH $h_i$ (respectively, an MSS $s_j$) bcm-Hbroadcasts (respectively, bcm-Sbroadcasts) a message $m$, it bcm-Hdelivers (respectively, bcm-Sdelivers) $m$.
	\item  \textit{BCM-Termination 2}:  If an MSS node $s_j$ bcm-Sdelivers a message $m$ from an MH $h_i$, then all MSS nodes bcm-Sdeliver the message $m$ from $h_i$ through $s_j$.  
	\item  \textit{BCM-Termination 3}:  If a non-faulty MH bcm-Hdelivers a message $m$ from an MH $h_i$, then all non-faulty MH nodes bcm-Hdeliver the message $m$ from $h_i$.
	\item  \textit{BCM-Termination 4}: If a message $m$ is bcm-Sbroadcast by an MSS $s_j$ after $h_i$ is disconnected from a group and the same message is bcm-Sbroadcast by an MSS $s_k$ before $h_i$ is connected in $s_k$, then $h_i$ will eventually bcm-Hdeliver the message $m$ from $s_k$. 
	\item  \textit{BCM-Causality 1}: If an MH $h_i$ bcm-Hbroadcasts a message $m_1$ when it is connected to an arbitrary group and subsequently bcm-Hbroadcasts another message $m_2$ after handoff to another group, then all non-faulty MH nodes in all groups (respectively, all MSS nodes) bcm-Hdeliver (respectively, bcm-Sdeliver) $m_1$ before $m_2$. 
	\item  \textit{BCM-Causality 2}: If an MSS $s_j$ bcm-Sbroadcasts a message $m_1$ in an arbitrary group after the MH $h_i$ has been disconnected, and then another MSS $s_k$ bcm-Sdelivers $m_1$, and simultaneously, when $h_i$ connects to $s_k$, $s_k$ bcm-Sbroadcasts message $m_2$ in its group, $h_i$ first bcm-Hdelivers $m_1$ and then $m_2$.
	\item  \textit{BCM-Causality 3}: If an MH node (respectively, an MSS node) first bcm-Hdelivers or bcm-Hbroadcasts (respectively, bcm-Sdelivers or bcm-Sbroadcasts) a message $m_1$ and then bcm-Hbroadcasts (respectively, bcm-Sbroadcasts) the message $m_2$, no non-faulty MH node (respectively, no MSS node) bcm-Hdelivers (respectively, bcm-Sdelivers) $m_2$ before $m_1$.
	\item  \textit{BCM-Safety}: If a Byzantine MH $h_i$ bcm-Hbroadcasts a message $m$ in the system, no MSS node (respectively, no non-faulty MH node) bcm-Sdelivers (respectively, bcm-Hdelivers) $m$.
\end{enumerate}

\begin{figure}
    \centering
    \includegraphics[width=1\linewidth]{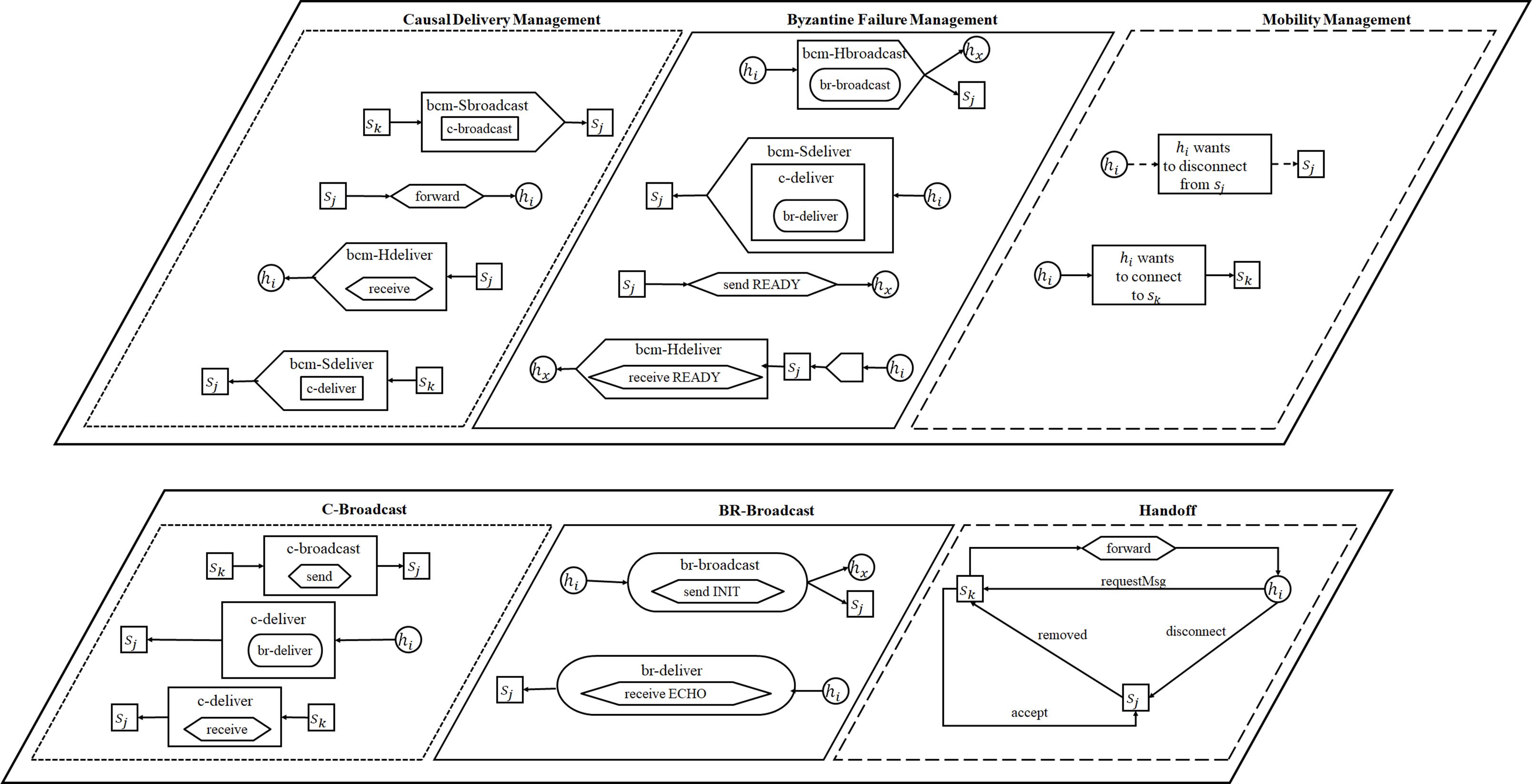}
    \caption{The relationship between the operations in different layers}
    \label{fig5}
\end{figure}

The properties mentioned above, BCM-Causality 1 and BCM-Causality 2 state that even handoff of an MH does not create a disturbance in the causal order of messages. The property BCM-Causality 3 states that all non-faulty MH and MSS nodes eventually have a causal agreement between their delivered messages. Figure \ref{fig5} illustrates the three mentioned layers and the way of communication between MSS and MH nodes in these layers. According to Fig. \ref{fig5}, the BCM-Broadcast algorithm is built on top of the BR-Broadcast, C-Broadcast, and Handoff algorithms. Thus, the handoff is executed independently of BR-Broadcast and C-Broadcast, but the C-Broadcast is implemented on top of the BR-Broadcast.
\section{BCM-Broadcast: The Proposed Algorithm}\label{sec:alg}
This section describes the BCM-Broadcast layered algorithm, as the Byzantine failure management layer uses the BR-Broadcast algorithm, the causal delivery management layer uses the C-Broadcast algorithm, and the mobility management layer uses the Handoff algorithm. Section \ref{sec4.1} introduces the data structures of MH and MSS nodes, and Section \ref{sec4.2} presents the pseudo-code of the proposed algorithm separately for different layers. In addition, Section \ref{sec4.3} describes a scenario of implementing the BCM-Broadcast algorithm in all layers.

\subsection{Data Structures}\label{sec4.1}
In this section, we describe the variables used in the algorithm separately for $h_i$ and $s_j$. 
\begin{itemize}
    \item 	Local data structure for an MH $h_i$:
    \begin{itemize}
       \item $TELEPOINT_i$: The id of the MSS to which $h_i$ is connected. If $h_i$ is not connected to any MSS, it takes a \textit{null} value.
	   \item $seq_i$: the $seq_i$ holds the sequence number of the last message that $h_i$ has broadcast. The sequence number of each message is a unique value.
	   \item $H_i\_{DELIV[N_{mss}]}$: a vector of natural numbers, where \linebreak $H_i\_{DELIV[s_k]}$ contains the number of messages $h_i$ has delivered from $s_k$.
	   \item $VIEW_i$: the set of MH nodes with which $h_i$ can exchange messages where $h_i\in VIEW_i$.
	\item $NVIEW_i = |VIEW_i|$: the number of MH nodes that are currently in the group of $h_i$, and $h_i$ can exchange messages with them.
	\item $H_i\_DELIVfromH[NVIEW_i]$: a vector of natural numbers, where $H_i\_DELIVfromH[h_x]$ contains the number of messages $h_i$ has delivered from $h_x$. 
	\item $LAST\_BCAST_i$:  a variable that holds the last broadcast message by $h_i$. 
    \end{itemize}
    \item   Local data structure for an MSS $s_j$ :
    \begin{itemize}
    \item $MH_j$: a set of MH nodes connected to an arbitrary MSS $s_j$.
	\item $NMH_j = |MH_j|$: the number of MH nodes that are connected to the MSS $s_j$.
	\item  $sn_j$: the sequence number of the last message delivered from MH nodes connected to $s_j$.
	\item $cb_j$: the information of the delivered messages is attached as a causal barrier to the broadcast message and its initial value $\phi$.
	\item $MOVING_j$: the set of MH nodes leaving the group of $s_j$ whose connection in the destination group has not yet been confirmed to $s_j$. 
\item $CONNECTING_j$: the set of MH nodes connecting to $s_j$ for which the handoff information has not yet been received by $s_j$ from their source MSS.
	\item $S_j\_RECEIVEDfromS$: a buffer for broadcast messages that have reached MSS $s_j$ from an MSS node but have not yet been bcm-Sdelivered.
	\item $S_j\_RECEIVEDfromH$: a buffer for broadcast messages that have reached MSS $s_j$ from an MH node but have not yet been bcm-Sdelivered.
    \item $S_j\_DELIV[N_{mss}]$: a vector of natural numbers, where\linebreak $S_j\_DELIV[s_k]$ contains the number of messages $s_j$ has delivered from $s_k$.
    \item $S_j\_KnowBcast[NMH_j]$: a vector of natural numbers, where \linebreak $S_j\_KnowBcast[i]$  contains the number of messages $h_i$ has broadcast and $s_j$ is aware of it.
	\item $DELIV_{j}[NMH_j]$: a vector of natural numbers, where $DELIV_{j}[h_i]$ contains the number of messages $s_j$ has delivered from $h_i$.
	\item $DELIV\_MES_j$: a queue of the last messages that $s_j$ delivered from other MSS nodes.
    \item $FORWARDED_j[N_{mss}]$: a vector of natural numbers, where \linebreak $FORWARDED_j[s_k]$ contains the number of messages forwarded from $s_k$ to the MH nodes of $s_j$.
    \item $RECV\_FORWARD_j[N_{mss}]$: a matrix where \linebreak $RECV\_FORWARD_j[s_k]$ contains $FORWARDED_k[N_{mss}]$.
	\item $S_j\_VIEW$: the set of MSS nodes with which $s_j$ can exchange messages, where $s_j\in S_j\_VIEW$.
 \item $MIN\_{F}[MSS]$: a natural number, where $MIN\_{F}[s_k]$ contains the minimum number of messages that different MSS nodes have broadcast in their group from $s_k$.
    \end{itemize}
\end{itemize}

\subsection{BCM-Broadcast Layers}\label{sec4.2} 
This section presents the algorithm associated with each layer separately. Section \ref{sec4.2.1} describes the Byzantine Reliable Broadcast algorithm for the Byzantine failure management layer; Section \ref{sec4.2.2} describes the Causal Broadcast algorithm for the causal delivery management layer. Also, Section \ref{sec4.2.3} describes the Handoff algorithm for the mobility management layer, and Section \ref{sec4.2.4} presents the BCM-Broadcast algorithm. 

\subsubsection{Byzantine Failure Management}\label{sec4.2.1} 
Algorithms \ref{alg.1} and \ref{alg.2} are revised versions of the BR-Broadcast algorithm presented in \cite{2}. The BR-Broadcast algorithm \cite{2} has also been inspired by Bracha’s BR-Broadcast algorithm \cite{3}. The BR-Broadcast algorithms \cite{2, 3} are presented for an arbitrary process $P_i$ and we revise it for two nodes $h_i$ and $s_j$. Bracha \textit{et al}. \cite{3} have used three protocol messages, \textit{INIT()}, \textit{ECHO()}, and \textit{READY()}, in their algorithm to ensure agreement between processes for the delivery of identical messages in the presence of Byzantine processes. A message \textit{INIT()} carries an application message, while the messages \textit{ECHO()} and \textit{READY()}, carry a process identity and an application message and are exchanged between processes to confirm an application message. An arbitrary process needs to exchange $2n^2-n-1$ protocol messages to deliver an application message, where $n$ is the number of all processes. Since we have considered an MSS node as a non-faulty stationary node, each MSS plays a significant role in reaching a collective consensus in a group. Also, we consider the energy of MH nodes to be limited and they should exchange fewer messages. For this reason, we present a new approach where processes need less message exchange to confirm the delivery of a message and reach an agreement on delivering identical messages. We also use three protocol messages, \textit{INIT()}, \textit{ECHO()}, and \textit{READY()} that are exchanged between MH nodes and MSS node in a group. When the group members (non-faulty or Byzantine MH nodes and MSS node) receive the message \textit{INIT()}  from an MH (non-faulty or Byzantine), they send the message \textit{ECHO()}  only to the MSS node instead of sending it to all MH nodes in the group. As a reliable station, the MSS node is assumed to be a final decision-maker. Since the MSS must guarantee the causal order of the messages in the causal delivery management layer, when it receives the message \textit{ECHO()} from $\frac{2NMH_j}{3} + 1$  number of MH nodes in the group, the MSS br-delivers the message, but it does not send \textit{READY()} to the MH nodes at this time.
\setstretch{1.35}
\begin{algorithm}[hbt!]
\caption{Revised version of Bracha’s BR-Broadcast algorithm \cite{2} (code for $h_i$) }\label{alg.1}
\begin{algorithmic}[1]
\Statex  \textbf{Operation} br-broadcast$(m,seq_i,h_i)$ at $h_i$ is
\State Send \textit{INIT}$(m, seq_i, h_i)$ to $VIEW_i$ and $TELEPOINT_i$;
\State $LAST\_BCAST_{i} = m$;
\Statex \textbf{When} a message \textit{INIT}$(m, seq_i, h_i)$ is received from $h_i$

\If {$seq_i = H_i\_DELIVfromH[i]$}
\State {disregard; {\small \color{blue} //$m$ is a duplicate message}}
\Else
\State {Send \textit{ECHO}$(<i,seq_i>, m)$ to $TELEPOINT_i$; }	
\EndIf.
\end{algorithmic}
\end{algorithm}

Algorithm \ref{alg.1} is executed for an MH $h_i$. When $h_i$ invokes \textit{br-broadcast()} in order to broadcast a message $m$, $h_i$ sends $m$ in the form of a message \textit{INIT()} to all MH nodes in $VIEW_i$ and the MSS node, $TELEPOINT_i$ (Line 1). Then, $h_i$ temporarily stores $m$ as the last broadcast message in $LAST\_BCAST_i$ (Line 2). When an MH $h_i$ receives a message \textit{INIT()}, it disregards $m$ if it is a duplicate message (Line 4); otherwise, it sends a message \textit{ECHO()} to the MSS in its group (Line 6).

According to Algorithm \ref{alg.2}, when an MSS $s_j$ receives a message \textit{INIT()} from an MH $h_i$ (non-faulty or Byzantine), if $seq_i$ of this message is greater than the number of the last broadcast message by $h_i$, then the message is not duplicated (Line 1). Thus, $s_j$  waits to receive the \textit{ECHO()} from the MH nodes of its group (Line 2). If the message is duplicated, $s_j$ disregards the received message (Line 4). When $s_j$ has received the \textit{ECHO()} from  $\frac{2NMH_j}{3} + 1$ different MH nodes, if all received messages \textit{ECHO()} have a common $seq_i$ and the same message, $s_j$ br-delivers $m$ (Line 10). Otherwise, $s_j$ disregards the messages \textit{INIT()} and \textit{ECHO()} in Line 8. Also, if $s_j$ does not receive this number of \textit{ECHO()}, then it ignores \textit{INIT()} in Line 13.

\begin{algorithm}[hbt!]
\caption{Revised version of Bracha’s BR-Broadcast algorithm \cite{2} (code for $s_j$) }\label{alg.2}
\begin{algorithmic}[1]
\Statex \textbf{When} a message \textit{INIT}$(m,seq_i,h_i)$ is received from $h_i$
\If{$seq_i > S_j\_KnowBcast[i]$}           
\State {Wait for \textit{ECHO}$(<i,seq_i>, m)$ from $MH_j$ {\small \color{blue} // $m$ is not a duplicate message} }
\Else 
\State  disregard;
\EndIf.
\Statex \textbf{When} a message \textit{ECHO}$(<i,seq_i>, m)$ is received from $h_i$
\If{\textit{ECHO}$(<i,seq_i>, m)$ received from strictly more than $\frac{2NMH_j}{3} + 1$ different $mh\in MH_j$    and not yet br-deliver} 
\If{$s_j$ received \textit{ECHO}$(<i,seq_{i}>, m)$ from $h_y$ and \textit{ECHO}$(<i,seq_{i}>, m^{'})$ from $h_x$}
\State {disregard \textit{INIT}$(m,seq_{i}, h_{i})$ and the messages \textit{ECHO}$(<i,seq_{i}>,-)$};  
\Else      
\State br-deliver $(m,seq_{i}, h_i)$; 
\EndIf
\Else 
\State {disregard the \textit{INIT}$(m,seq_{i}, h_{i})$;  }
\EndIf.
\end{algorithmic}
\end{algorithm}

\begin{algorithm}
\caption{C-Broadcast algorithm (Revised version of the algorithms \cite{1, 2}) - Part 1}\label{alg.3.p1}
\begin{algorithmic}[1]
\Statex \textbf{When} $s_j$ br-delivers a message $(m,seq_i, h_i)$ from $h_i$
\If{$seq_i = S_j\_KnowBcast[i] + 1$} {\small \color{blue} //if $s_j$ has br-delivered all messages preceding $m$ from $h_i$}
\State c-deliver $(m,h_i)$;   
\Else 
\State $S_j\_RECEIVEDfromH = S_j\_RECEIVEDfromH\bigcup \left\{(m, h_i, seq_i)\right\}$; 
\EndIf.
\Statex \textbf{When} $s_j$ wants to c-broadcast $(m)$ 
\State	  $sn_j = sn_{j} + 1$;
\State	send $(<m,s_j,sn_j>,cb(m), FORWARDED_j)$ to $S_j\_VIEW$;
\State $cb_j = \phi$;
\Statex \textbf{When} $s_k$ receives a message $(<m,s_j,sn_j>,cb(m), FORWARDED_j)$ from $s_j$, where $s_k$ can be the same as $s_j$
\If{{$sn_j = S_{k}\_DELIV[s_j] + 1$} and {$\forall {(l, sn^{'})}\in {cb(m)}:{S_k\_DELIV[s_l]\geq {sn^{'}}}$}} {\small \color{blue} //if $s_k$ has bcm-Sdelivered all messages preceding $m$ from $s_j$ and the messages within $cb$ that have arrived along with $m$, then $s_k$ c-delivers $m$}  
\State $cb_k = cb_k\setminus {cb(m)}\bigcup \left\{(s_j, sn_j)\right\}$;
\State c-deliver $(m, s_j)$    
\State $S_k\_DELIV[s_j] = S_k\_DELIV[s_j] + 1$;
\State $DELIV\_MES_k = DELIV\_MES_k\bigcup \left\{(m, s_{j}, sn_{j} )\right\}$;
\State $RECV_FORWARD_k[s_j] = max(RECV_FORWARD_k[s_j], FORWARDED_j)$;
\For{\textbf{each}$(m^{'}, s^{'}, sn^{'})\in S_k\_RECEIVEDfromS$}    {\small \color{blue} //$s_k$ has previously received a message $m^{'}$ with sequence number $sn^{'}$ from $s^{'}$ but not yet c-delivered}
\If{$sn^{'} = S_k\_DELIV[s^{'}] + 1$} {\small \color{blue} //if $s_k$ has previously c-delivered all messages preceding $m^{'}$ from $s^{'}$, then $s_k$ c-delivers $m^{'}$ } 
\State delete $(m^{'}, s^{'}, sn^{'})$  from $S_k\_RECEIVEDfromS$ ;
\State c-deliver $(m^{'}, s^{'})$;    
\State $S_k\_DELIV[s^{'}] = S_k\_DELIV[s^{'}] + 1$;
\State {$DELIV\_MES_{k} = DELIV\_MES_{k}\bigcup \left\{(m^{'}, s^{'}, sn^{'})\right\}$};
\EndIf   
\EndFor
\For{\textbf{each}$(m^{''}, h_i, seq_{i})\in {S_k\_RECEIVEDfromH}$}     {\small \color{blue} //$s_k$ has previously br-delivered a message $m^{''}$ with sequence number $seq_i$ from $h_i$ but not yet c-delivered}
\algstore{bkbreak}
\end{algorithmic}
\end{algorithm}

\addtocounter{algorithm}{-1}
\begin{algorithm}[hbt!]
\caption{C-Broadcast algorithm (Revised version of the algorithms \cite{1, 2}) - Part 2}\label{alg.3.p2}
\begin{algorithmic}[1]
\algrestore{bkbreak}
\If{$seq_{i} = S_k\_KnowBcast[i] + 1$ }  {\small \color{blue} // if $s_k$ has previously bcm-Sdelivered all messages preceding $m^{''}$ from $h_i$,  then $s_k$ c-delivers $m^{''}$}
\State delete $(m^{''}, h_i, seq_{i})$  from $S_k\_RECEIVEDfromH$ ;
\State c-deliver $(m^{''}, h_{i})$;
\State $S_k\_DELIV[s_{k}] = S_k\_DELIV[s_{k}] + 1$;
\EndIf   
\EndFor
\State 	$MIN\_{F}[MSS] = min(RECV\_FORWARD_{k}[s]:\forall s\in {MSS})$;     {\small \color{blue} // the minimum number of messages that different MSS nodes have broadcast in their group from other MSS nodes}
\For{\textbf{each}$(m^{'}, s^{'}, sn^{'})\in{DELIV\_MES_k}$} 	 {\small \color{blue} // $s_k$ has previously bcm-Sdelivered a message $m^{'}$ with sequence number $sn^{'}$ from $s^{'}$ and has stored in $DELIV\_MES_k$}
\If{$MIN\_F[s^{'}]\geq sn^{'}$ and $\forall (h_i, H_{i}\_DELIV[s^{'}])\in {CONNECTING_k}: H_i\_DELIV[s^{'}]\geq sn^{'}$}  {\small \color{blue} // If all MSS nodes have forwarded $m^{'}$ or even the messages following $m^{'}$ to the MH nodes of their group, and  $h_i$, which is connecting to $s_k$, has delivered all messages preceding $m^{'}$ or even $m^{'}$ in the source group,  then $s_k$ removes $m^{'}$  from $DELIV\_MES_k$}
\State 	delete $(s^{'}, sn^{'}, m^{'}$)  from $DELIV\_MES_k$;    \EndIf
\EndFor
\Else
\State $S_k\_RECEIVEDfromS = S_k\_RECEIVEDfromS \bigcup \left\{(m, s_{j},sn_{j})\right\}$;   
\EndIf.
\end{algorithmic}
\end{algorithm}

\subsubsection{Causal Delivery Management}\label{sec4.2.2}
In this layer, an MSS $s_j$ c-delivers the messages it has br-delivered in the Byzantine failure management layer, and $s_j$ c-broadcasts the messages it has bcm-Sdelivered from an MSS $s_k$. Algorithm \ref{alg.3.p1} presents how $s_j$ broadcasts and delivers the messages in a causal order. This algorithm is inspired by the algorithms presented in \cite{1, 2}. Lines 6 to 12 are revised versions of Lines 1 to 8 of the Byzantine Causal Order (BCO) Broadcast algorithm \cite{2}, which guarantees the causal broadcast and delivery of messages by an MSS node. Moreover, Lines 30 to 33 are revised versions of the algorithm presented in \cite{1}, which removes delivered messages forwarded to all MH nodes from the list $DELIV\_MES$.

When $s_j$ br-delivers the message $m$ in the Byzantine failure management layer from $h_i$, it first checks whether the sequence number $seq_i$ of this message is only a unit greater than the last message broadcast by $h_i$ (Line 1). In this case, $s_j$ has bcm-Sdelivered all the previous messages. Thus, $s_j$ c-delivers this message (Line 2); otherwise, it adds $m$ to the buffer of received messages to c-deliver it after bcm-Sdelivered the previous messages (Line 4). When $s_j$ intends to c-broadcast a locally delivered message to other MSS nodes, first, it increases the global sequence number $sn_j$ (Line 6). Then it attaches the causal barrier and $FORWARDED_j$ to the end of $m$ and afterward sends it to all MSS nodes (Line 7). Finally, it empties the content of the causal barrier (Line 8).

When an MSS $s_k$ receives a global message from an arbitrary MSS $s_j$, it first checks whether all the messages that have been causally broadcast before this message have been delivered by $s_k$ (Line 9). If $s_k$ has delivered previous messages, it updates the content of its causal barrier (Line 10) and then c-delivers $m$ at the causal delivery management layer (Line 11). Then, in Lines 12 to 14 it updates the local variables. $s_k$ checks the contents of the buffer of $S_k\_RECEIVEDfromS$. Suppose a message $m^{'}$ waits for $s_k$ to deliver a message $m$ before $m^{'}$. Thus, if $s_k$ has delivered the message $m$, it removes those from the buffer of $S_k\_RECEIVEDfromS$ and c-delivers it (Lines 15 to 18). Then, $s_k$ increases the number of messages it has delivered from $s^{'}$ (Line 19). Also, $s_k$ adds a copy of the delivered message to the queue of $DELIV\_MES_k$ (Line 20). $s_k$ checks the contents of the buffer of $S_k\_RECEIVEDfromH$ whether there are the messages that have been waiting to bcm-Sdeliver by $s_k$, $s_k$ removes those from the buffer of $S_k\_RECEIVEDfromH$ and c-delivers it (Lines 23 to 26). Then, in Line 27 it updates the local variables. Since the permanent maintenance of copies of delivered messages in the queue $DELIV\_MES_k$ requires a lot of storage space, $s_k$ eliminates the messages delivered in all groups from this queue. $s_k$ stores the minimum number of messages that different MSS nodes have broadcast in its group from other MSS nodes in the variable $MIN\_F$ (Line 30). $s_k$ checks the contents of the queue of $DELIV\_MES_k$, if a message $m^{'}$ with sequence number $sn^{'}$ from $s^{'}$ exists, and the message or messages after $m^{'}$ have been forwarded to all groups (Lines 31 and 32). Then, $s_k$ eliminates $m^{'}$ from $DELIV\_MES_k$ (Line 33). If $s_k$ has not yet received the previous messages, it temporarily stores $m$ in the buffer of $S_{k}\_RECEIVEDfromS$ (Line 37).

\subsubsection{Mobility Management}\label{sec4.2.3}
Algorithms \ref{alg.4} and \ref{alg.5} are the revised version of the handoff algorithm presented in \cite{1}. In Algorithm \cite{1}, an MH node first sends a connect message to join an MSS node $s_k$ and then joins $s_k$ after being confirmed by $s_k$. By contrast, in our algorithm, when an MH node reaches the radio range of an MSS node, it automatically joins that MSS node. Then, it sends a request message to deliver the messages in the queue of $DELIV\_MES$ of the MSS node. Algorithm \ref{alg.4} presents the operations that an MH $h_i$ performs to handoff from an MSS $s_j$ to another MSS $s_k$. When $h_i$ wants to leave a group and disconnects from $s_j$, first it sends the message \textit{disconnect()} attached with the destination id and the variable $H_i\_DELIV$ to $s_j$, where $H_i\_DELIV$ is an array of numbers of messages $h_i$ has delivered from different MSS nodes (Line 1). Then, it sets its connection point and view to null (Lines 2 and 3). Afterwards, $h_i$ waits until it reaches the radio range of $s_k$ (Line 4). After detecting the radio range of $s_k$, $h_i$ sets its \textit{TELEPOINT} to $s_k$ (Line 5), then sends the \textit{requestMsg()} as a request message to $s_k$ (Line 6).

\begin{algorithm}[hbt!]
\caption{Revised version of the handoff algorithm in \cite{1} (code for $h_i$)}\label{alg.4}
\begin{algorithmic}[1]
\Statex \textbf{When} $h_i$ wants to disconnect from $s_j$
\State Send disconnect $(s_k, H_i\_DELIV)$ to $s_j$
\State $TELEPOINT_i = null$;
\State $VIEW_i = null$;
\State Wait (until the radio range of $s_k$ is detected);
\Statex \textbf{When} $h_i$ wants to connect to $s_k$
\State $TELEPOINT_i = s_k$;
\State Send requestMsg$(s_j, H_i\_DELIV)$ to $s_k$
\end{algorithmic}
\end{algorithm}
Algorithm \ref{alg.5} describes the operation a source/destination MSS performs upon receiving a \textit{disconnect()}/\textit{requestMsg()}. When $s_j$ receives a message \textit{disconnect()} from $h_i$, first it checks whether $h_i$ is in $MH_j$ (Line 1). If $h_i$ is still in $MH_j$, $s_j$ removes it from $MH_j$ and reduces the number of processes connected to $s_j$ (Lines 2 and 3), then it temporarily stores $h_i$ in the set $MOVING_j$ (Line 4). Finally, $s_j$ sends information about $h_i$ to the destination MSS in the form of a message \textit{removed()} (Line 5). If $h_i$ has previously left the group and is not in $MH_j$, $s_j$ disregards the message \textit{disconnect()} (Line 7).

\begin{algorithm}[hbt!]
\caption{Revised version of the handoff algorithm in \cite{1}}\label{alg.5}
\begin{algorithmic}[1]
\Statex \textbf{When} $s_j$ receives disconnect$(s_k, H_i\_DELIV)$ from $h_i$
\If{ $h_{i}\in {MH_j}$}
\State $MH_{j} = MH_{j} - \left\{h_{i}\right\}$;
\State $NMH_j = NMH_{j} - 1$;
\State $MOVING_j = MOVING_{j}\bigcup \left\{(h_i, H_i\_DELIV)\right\}$;
\State Send removed $(s_j, h_i, H_i\_DELIV, S_j\_KnowBcast[i])$  to $s_k$
\Else
\State disregard;
\EndIf.
\Statex \textbf{When} $s_k$ receives requestMsg $(s_j, H_i\_DELIV)$  from $h_i$
\If{  $h_{i}\notin {MH_{k}}$  and $h_{i}\notin {CONNECTING_k}$ }   
\State $CONNECTING_k = CONNECTING_k\bigcup \left\{(h_i, H_i\_DELIV)\right\}$;
\Else
\State disregard;
\EndIf.
\Statex \textbf{When} $s_k$ receives removed$(s_j, h_i, H_i\_DELIV,S_j\_KnowBcast[i])$  from $s_j$
\State $MH_{k} = MH_{k}\bigcup \left\{h_i\right\}$;
\State $NMH_{k} = NMH_{k} + 1$;
\State $S_k\_KnowBcast[i] = S_j\_KnowBcast[i]$;
\State Send $(s_{j}, accept(h_{i}))$;
\State $CONNECTING_{k} = CONNECTING_{k} - \left\{(h_{i}, H_i\_DELIV)\right\}$;
\For{\textbf{each} $(s^{'}, sn^{'}, m^{'})\in {DELIV\_MES_k}$}  {\small \color{blue} // An arbitrary MSS node $s^{'}$ has previously broadcast the message $m^{'}$ with sequence number $sn^{'}$ and is now in $DELIV\_MES_k$.}
\If{$H_i\_DELIV[s^{'}] < sn^{'}$}     {\small \color{blue} //$h_i$ has not bcm-Hdeliver the message $m^{'}$}
\State forward $(m^{'}, s^{'}, MH_{k})$ to $h_i$ 
\EndIf
\EndFor    {\small \color{blue} //the for each loop is executed as many messages as there are in the $DELIV\_MES_k$.}
\Statex \textbf{When} $s_j$ receives$(s_{j}, accept(h_{i}))$ from $s_k$
\State $MOVING_{j} = MOVING_{j} - \left\{(h_{i}, H_i\_DELIV)\right\}$;
\end{algorithmic}
\end{algorithm}

When $s_k$ receives \textit{requestMsg()} from $h_i$, if $h_i$ is not in $MH_k$ or not in the list of \textit{CONNECTING}, meaning $h_i$ has not previously joined $s_k$ or is not in Handoffing state, then $s_k$ adds it to the list of \textit{CONNECTING} (Lines 9 and 10). Otherwise, it ignores the message of \textit{requestMsg()} (Line 12). When $s_k$ receives a message \textit{removed()} from $s_j$, it first adds $h_i$ to $MH_k$ (Line 14), then increases the total number of MH nodes connected to the system (Line 15) and $s_k$ updates its information about $h_i$ from $s_j$'s information (Line 16). Afterward, it sends the message \textit{accept()} to $s_j$ (Line 17) and removes $h_i$ from the list of \textit{CONNECTING} (Line 18). Finally, $s_k$ forwards to $h_i$ the messages that it has in the delivered messages queue and which have not yet been delivered by $h_i$ (Lines 19 to 23). When $s_j$ receives the message \textit{accept()} from $s_k$, it removes $h_i$ from the list of  \textit{MOVING} (Line 24).

\subsubsection{The BCM-Broadcast Algorithm}\label{sec4.2.4}
Algorithm \ref{alg.6} introduces the BCM-Broadcast algorithm by an MSS node. In Algorithm \ref{alg.6}, when $s_j$ wants to broadcast the message $m$ to other MSS nodes at the application level, it invokes the operation \textit{c-broadcast()} from the causal delivery management layer (Line 1). In second "When" of Algorithm \ref{alg.2}, when $s_j$ receives \textit{ECHO($m$)} from non-faulty MH nodes and br-delivers $m$ in Line 10 of Algorithm \ref{alg.2}, it does not send \textit{READY($m$)} until it causally delivers $m$. Thus, $s_j$ in Line 2 of Algorithm \ref{alg.3.p1} c-delivers $m$ from $h_i$. Finally, in Line 2 of algorithm \ref{alg.6}, it bcm-Sdelivers. Then, $s_j$ sends \textit{READY($m$)} to all MH nodes in its radio range (Line 5 of Algorithm \ref{alg.6}). Afterwards, $s_j$ updates the local variable $FORWARDED_j$. In addition, by invoking \textit{c-broadcast()} from the causal delivery management layer, it causally broadcasts $m$ to other MSS nodes (Line 9 of Algorithm \ref{alg.6}). When $s_k$ receives $m$ from $s_j$, first it bcm-Sdelivers $m$, then it forwards $m$ to all MH nodes in its group (Lines 10 to 13).

\begin{algorithm}[hbt!]
\caption{BCM-Broadcast algorithm (code for an MSS node) }\label{alg.6}
\begin{algorithmic}[1]
\Statex \textbf{When} $s_j$ wants to bcm-Sbroadcast($m$) 
\State c-broadcast ($m$);
\Statex \textbf{When} $s_j$ c-delivers a message $(m, h_{i})$ from $h_i$
\State bcm-Sdeliver($m$);
\State $S_j\_KnowBcast[i] = S_j\_KnowBcast[i] + 1$;
\For{\textbf{each} $mh\in {MH_j}$ }
\State send READY$(m, h_{i})$  to $MH_j$; 
\State $FORWARDED_k[s_{j}] = FORWARDED_k[s_{j}] + 1$;
\EndFor
\State c-broadcast($m$);
\Statex \textbf{When} $s_k$ c-delivers a message $(m, s_{j})$ from $s_j$   
\State bcm-Sdeliver($m$);
\If{$s_{k}\neq s_{j}$}  {\small \color{blue} //If $s_k$ has  bcm-Sdelivered the message $m$ from another MSS, $s_k$ forwards $m$ to the MH nodes of its group because $s_k$ has already broadcasts $m$ in its group as  \textit{READY$(m, h_{i})$}}
\For{\textbf{each} $mh\in {MH_k}$ }
\State forward $(m, s_{j})$  to $MH_k$;
\State $FORWARDED_k[s_{j}] = FORWARDED_k[s_{j}] + 1$;
\EndFor
\EndIf.
\end{algorithmic}
\end{algorithm}

\begin{algorithm}[!hbt]
\caption{BCM-Broadcast algorithm (code for $h_i$)}\label{alg.7}

\begin{algorithmic}[1]
\Statex \textbf{When} $h_i$ wants to bcm-Hbroadcast ($m$) 
\If{$m = LAST\_BCAST_i$ }   {\small \color{blue} // $m$ is a duplicate message}
\State $seq_{i} = seq_{i}$;
\Else
\State $seq_{i} = seq_{i} + 1$;
\EndIf
\State br-broadcast $(m, seq_{i}, h_{i})$ to $VIEW_i$ and $TELEPOINT_i$;
\Statex \textbf{When} $h_i$ receives a READY$(m, h_{x})$  from $TELEPOINT_i$
\State bcm-Hdeliver ($m$);
\State $H_i\_DELIV[s_{j}] = H_i\_DELIV[s_{j}] + 1$;
\State $H_i\_DELIVfromH[x] = H_i\_DELIVfromH[x] + 1$;
\Statex \textbf{When} $h_i$ receives a message $(m^{'},s^{'},MH_{k})$ from $s_k$ 
\State bcm-Hdeliver ($m^{'}$);
\State $H_i\_DELIV[s^{'}] = H_i\_DELIV[s^{'}] + 1$;
\State $VIEW_{i} = MH_k$;
\Statex \textbf{When} $h_i$ receives a message $(m, s_{j})$  from $s_k$
\State bcm-Hdeliver ($m$);
\State $H_i\_DELIV[s_{j}] = H_i\_DELIV[s_{j}] + 1$;
\end{algorithmic}
\end{algorithm}

Algorithm \ref{alg.7} introduces the steps of execution of the BCM-Broadcast algorithm by $h_i$. When $h_i$ wants to bcm-Hbroadcast a message $m$ to all MH and MSS nodes in the system at the application level, it first compares $m$ with the last message it broadcasts (Line 1). If the message is duplicated, then $h_i$ br-broadcasts $m$ to MH nodes and MSS node in its group with the same previous sequence number; Otherwise, it br-broadcasts $m$ with the new sequence number (Lines 2 to 6). The Byzantine failure management layer invokes the operation \textit{br-broadcast()} in Line 6 of Algorithm \ref{alg.7}. $s_j$ br-delivers the message $m$ in Line 10 of Algorithm \ref{alg.2} and then c-delivers it in Line 2 of Algorithm \ref{alg.3.p1} and afterwards, bcm-Sdelivers it in Line 2 of Algorithm \ref{alg.6} and finally in Line 5 of Algorithm \ref{alg.6} sends \textit{READY($m$)} to the MH nodes of its group. Thus, when $h_i$ receives \textit{READY($m$)} from $s_j$, it first bcm-Hdelivers it and then updates the local variables (Lines 7 to 9 of Algorithm \ref{alg.7}). In Line 21 of Algorithm \ref{alg.5}, when $h_i$ is being handed off to $s_k$, $s_k$ forwards the messages from the $DELIVE\_MES_k$ queue to $h_i$. When $h_i$ receives these messages from $s_k$, it first bcm-Hdelivers it and then updates the local variables (Lines 10 to 12 of Algorithm \ref{alg.7}). When $s_k$ bcm-Sdelivers the message $m$ from $s_j$ (Line 9 of Algorithm \ref{alg.6}), it forwards $m$ to the MH nodes in its group (Lines 10 to 13 of Algorithm \ref{alg.6}). When $h_i$ receives the global message $m$ from $s_j$ through $s_k$, it first bcm-Hdelivers $m$ and then updates the local variables (Lines 13 to 14 of Algorithm \ref{alg.7}).

\subsection{An Execution Scenario}\label{sec4.3}
\begin{figure}
    \centering
    \includegraphics[width=1\linewidth]{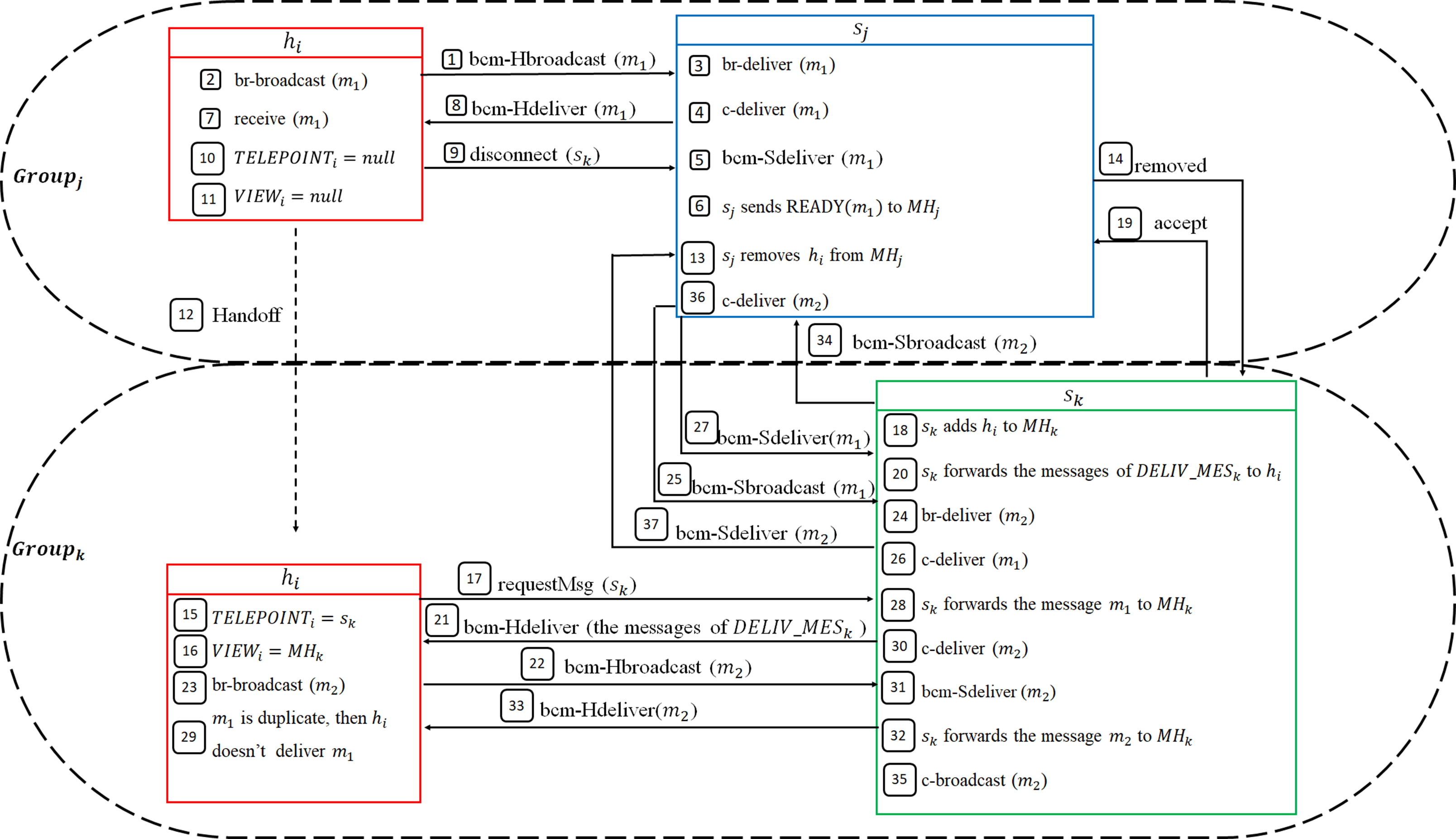}
    \caption{A scenario of execution of the BCM-Broadcast algorithm}
    \label{fig6}
\end{figure}

Figure \ref{fig6} illustrates an example execution  BCM-Broadcast algorithm; this example as a scenario contains 37 steps, which include: 1) An arbitrary MH $h_i$ bcm-Hbroadcasts a message $m_1$, 2) $h_i$ br-broadcasts $m_1$ in the Byzantine failure management layer, 3) $s_j$ receives $m_1$ and waits to receive the messages \textit{ECHO()} from the MH nodes of its group; thus, after receiving the message \textit{ECHO()} from $\frac{2NMH_j}{3}+1$ different MH nodes, it finally br-delivers $m_1$, 4) $s_j$ c-delivers $m_1$ in the causal delivery management layer, 5) $s_j$ bcm-Sdelivers $m_1$, 6) $s_j$ sends $m_1$ as a message \textit{READY()} to the MH nodes of its group, 7)  $h_i$ receives $m_1$, 8) $h_i$ bcm-Hdelivers $m_1$, 9) $h_i$ sends the messages disconnect to $s_j$, 10) $h_i$ sets its \textit{TELEPOINT} value to null, 11) $h_i$ sets its \textit{VIEW} value to null, 12) $h_i$ leaves $s_j$ and joins $s_k$, 13) $s_j$ removes $h_i$ from $MH_i$, 14) $s_j$ sends information about $h_i$ to $s_k$ by sending the message \textit{removed}, 15) $h_i$ sets its \textit{TELEPOINT} value to $s_k$ after reaching the radio range of $s_k$, 16) $h_i$ sets its \textit{VIEW} value to $MH_k$, 17) $h_i$ sends the message \textit{requestMsg()} to $s_k$, 18) $s_k$ adds $h_i$ to $MH_k$, 19) $s_k$ sends the message \textit{accept} to $s_j$, 20) $s_k$ forwards the messages in the list $DELIV\_MES_k$ to $h_i$, 21) $h_i$ bcm-Hdelivers the messages in the list $DELIV\_MES_k$, 22) $h_i$ bcm-Hbroadcasts a message $m_2$ in its group, 23) $h_i$ br-broadcasts $m_2$ in the Byzantine failure management layer. 

When $h_i$ invokes \textit{br-broadcast()} to broadcast $m_2$, $h_i$ sends $m_2$ as a message \textit{INIT($m_2$)} to all MH nodes in its group and $s_k$.  When the MH nodes receive a message \textit{INIT($m_2$)}, they send a message \textit{ECHO()} to $s_k$. If $s_k$ has received the \textit{ECHO()} from  $\frac{2NMH_k}{3}+1$ different MH nodes, which all received messages \textit{ECHO()} have a common $seq_i$ and the same message, then $s_k$ br-delivers $m_2$ (Step 24). When $h_i$ joined $s_k$, $s_k$ received the $S_{j}\_KnowBcast[i]$ value from $s_j$. Thus, by checking the $seq_i$ of $m_2$ and comparing it with $S_{k}\_KnowBcast[i]$, $s_k$ knows that it has not yet delivered $m_1$, then it temporarily stores $m_2$ in $S_{k}\_RECEIVEDfromH$. 25) $s_j$ bcm-Sbroadcasts $m_1$, 26) $s_k$ c-delivers $m_1$ in the causal delivery management layer, 27) $s_k$ bcm-Sdelivers $m_1$, 28) $s_k$ forwards the messages $m_1$ to all MH nodes in its group, 29) Since $h_i$ has already delivered the message $m_1$ from $s_j$, $h_i$ does not deliver $m_1$ again, 30) before delivering $m_1$, $s_k$ c-delivers $m_2$ from the list $S_{k}\_RECEIVEDfromH$, 31) $s_k$ bcm-Sdelivers $m_2$, 32) $s_k$ forwards the messages $m_2$ to all MH nodes in its group, 33) $h_i$ bcm-Hdelivers $m_2$, 34) $s_k$ bcm-Sbroadcasts $m_2$ to $S_{k}\_VIEW$, 35) $s_k$ c-broadcasts $m_2$ in the causal delivery management layer, 36) $s_j$ c-delivers $m_2$, and 37) $s_j$ bcm-Sdelivers $m_2$. This scenario shows that although $h_i$ broadcasts the message $m_1$ in one group and $m_2$ in another, all MSS and MH nodes participating in the algorithm deliver $m_1$ before delivering $m_2$.
\section{Correctness of the BCM-Broadcast Algorithm }\label{sec:corecct}
This section proves the correctness of the BCM-Broadcast algorithm, and analyzes its message complexity.
\begin{Lma} \label{lm.5}
    \normalfont If an MSS $s_j$ bcm-Sbroadcasts a message $m$, it bcm-Sdelivers $m$.
\end{Lma}    
\noindent \textbf{Proof.} When $s_j$ bcm-Sbroadcasts an arbitrary message $m$, according to Line 1 of Algorithm \ref{alg.6}, it invokes the operation c-broadcast() from the causal delivery management layer. Moreover, when $s_j$ intends to c-broadcast $m$, it sends the message $m$ to $S_{j}\_VIEW$ (Line 7 of Algorithm \ref{alg.3.p1}). According to the definition of $S_{j}\_VIEW$, $s_j$ itself exists in $S_{j}\_VIEW$. Based on the third "When" of Algorithm \ref{alg.3.p1}, $s_k$ can be the same as $s_j$. Thus, $s_k$ c-delivers $m$ (Line 11 of Algorithm \ref{alg.3.p1}), and then bcm-Sdelivers $m$ in Line 9 of Algorithm \ref{alg.6}. \qed

\begin{Theo}\label{theo.5.1}
\normalfont The BCM-Broadcast algorithm satisfies the property BCM-Validity 1.
\end{Theo} 
\noindent \textbf{Proof.} According to Line 9 of Algorithm \ref{alg.6}, when an arbitrary MSS $s_k$ bcm-Sdelivers the message $m$ from $s_j$, it has already c-delivered $m$ in the causal delivery management layer (according to Line 11 of Algorithm \ref{alg.3.p1}). Also, $s_k$ has previously received $m$ as a protocol message in the third "When" of Algorithm \ref{alg.3.p1}. When $s_k$ c-delivers $m$ from $s_j$, then $s_j$ has already c-broadcast $m$ (according to Lines 6 to 8 of Algorithm \ref{alg.3.p1}); also, $s_j$ has bcm-Sbroadcast $m$ before c-broadcasting it (Line 1 of Algorithm \ref{alg.6}). \qed    
\begin{Theo} \label{theo.5.2}
    \normalfont The BCM-Broadcast algorithm satisfies the property BCM-Validity 2.
\end{Theo}
\noindent \textbf{Proof.} An MH $h_i$ bcm-Hdelivers an arbitrary message $m$ in three ways from an MSS $s_j$, which include: 
\begin{enumerate}  
	\item According to Line 7 of Algorithm \ref{alg.7}, when $h_i$ bcm-Hdelivers the message $m$, then $s_j$ has previously sent the message $m$ as a message \textit{READY()} to $h_i$, according to Line 5 of Algorithm \ref{alg.6}.
	\item According to Line 10 of Algorithm \ref{alg.7}, when $h_i$ bcm-Hdelivers the message $m$, then after joining $h_i$ to $s_j$, $s_j$ has forwarded $m$ to $h_i$, according to Line 21 of Algorithm \ref{alg.5}.
    \item According to Line 13 of Algorithm \ref{alg.7}, when $h_i$ bcm-Hdelivers the message $m$, then $s_j$ has forwarded $m$ to $h_i$, according to line 12 of Algorithm \ref{alg.6}. 	\qed
\end{enumerate}
\begin{Lma}\label{lm.1}
    \normalfont If an MSS $s_j$ bcm-Sdelivers a message $m$ from an MH $h_i$, then $h_i$ bcm-Hbroadcast $m$.    
\end{Lma}
\noindent \textbf{Proof.} According to Line 2 of Algorithm \ref{alg.6}, when an arbitrary MSS $s_j$ bcm-Sdelivers the message $m$ from $h_i$, it has previously c-delivered $m$ in the causal delivery management layer (Line 2 of Algorithm \ref{alg.3.p1}). Also, $s_j$ has previously br-delivered $m$ in the Byzantine failure management layer, according to Line 10 of Algorithm \ref{alg.2}. When $s_j$ br-delivers $m$, then it has received $m$ as a message \textit{INIT(}) from $h_i$ (the first “When” of Algorithm \ref{alg.2}); also, $s_j$ received the \textit{ECHO()} from at least  $\frac{2NMH_{k}}{3}+1$ non-faulty MH nodes in its group (Line 6 of Algorithm \ref{alg.2}). When $s_j$ receives a message \textit{INIT()} from $h_i$, then $h_i$ has previously br-broadcast $m$ as a message \textit{INIT()} to $s_j$ and other MH nodes in its group in the Byzantine failure management layer (Line 1 of Algorithm \ref{alg.1}). Also, $h_i$ bcm-Hbroadcasts $m$ according to the first “When” of algorithm \ref{alg.7}.\qed

\begin{Lma} \label{lm.2}
    \normalfont If a non-faulty MH node bcm-Hdelivers a message $m$ from an MH $h_i$, then $h_i$ bcm-Hbroadcast $m$.
\end{Lma}
\noindent \textbf{Proof.} We know that an arbitrary MH $h_x$ bcm-Hdelivers the broadcast messages in the system through the MSS $s_j$ to which it is connected. Based on Lemma \ref{lm.1}, if $s_j$ bcm-Sdelivers $m$ from an MH $h_i$, then $h_i$ bcm-Hbroadcast $m$. Also, based on Theorem \ref{theo.5.2}, if a non-faulty MH $h_x$ bcm-Hdelivers a message $m$ from $s_j$, $s_j$ has forwarded $m$ to $h_x$. Thus, if $h_x$ bcm-Hdelivers $m$ from $h_i$, then $h_i$ bcm-Hbroadcast $m$. \qed

\begin{Theo} \label{theo.5.3}
    \normalfont The BCM-Broadcast algorithm satisfies the property BCM-Validity 3.
\end{Theo}
\noindent \textbf{Proof.} This Theorem follows directly from Lemma \ref{lm.1} and Lemma \ref{lm.2}.\qed

\begin{Lma} \label{lm.3}
    \normalfont An MSS node bcm-Sdelivers a message $m$ at most once.
\end{Lma}
\noindent \textbf{Proof.} When an arbitrary MSS $s_j$ receives a message $m$ as \textit{INIT()}, it checks whether it is duplicated in Line 1 of Algorithm \ref{alg.2}. If the message $m$ is duplicated, then $s_j$ disregards it, according to Line 4 of Algorithm \ref{alg.2}. Since the MSS nodes and their communication are reliable, no MSS node sends the same message more than once. Thus, no MSS node bcm-Sdeliver a message more than once. \qed

\begin{Lma} \label{lm.4}
    \normalfont A non-faulty MH node bcm-Hdelivers a message $m$ at most once.
\end{Lma}
\noindent \textbf{Proof.} We know that an arbitrary MH $h_i$ bcm-Hdelivers the broadcast messages in the system through the MSS $s_j$ to which it is connected. Also, based on Lemma \ref{lm.3}, $s_j$ bcm-Sdelivers a message $m$ at most once. Since the MSS nodes are non-faulty and reliable, those only forward a message $m$ to $h_i$ once. Thus, a non-faulty MH node bcm-Hdelivers a message $m$ at most once.  \qed        

\begin{Theo} \label{theo.5.4}
    \normalfont The BCM-Broadcast algorithm satisfies the property BCM-Integrity 1.
\end{Theo}
\noindent \textbf{Proof.} This theorem follows directly from Lemma \ref{lm.3} and Lemma \ref{lm.4}. \qed

\begin{Theo} \label{theo.5.5}
    \normalfont The BCM-Broadcast algorithm satisfies the property BCM-Integrity 2.
\end{Theo}
\noindent \textbf{Proof.} Consider that when $h_i$ is connected to $s_j$, $s_j$ bcm-Sbroadcasts an arbitrary message $m$. $h_i$ may face two situations: 1) before delivering $m$ from $s_j$, $h_i$ leaves $s_j$ and joins $s_k$. According to Algorithm \ref{alg.6}, when $s_k$ bcm-Sdelivers $m$ from $s_j$ (Line 9), $s_k$ forwards it to all MH nodes in its group (Line 12); thus, according to Line 13 of Algorithm \ref{alg.7}, $h_i$ bcm-Hdelivers $m$ from $s_k$ only once. 2) after delivering $m$ from $s_j$, $h_i$ leaves $s_j$ and joins $s_k$. When $h_i$ joins $s_k$, $s_k$ has already bcm-Sdelivered $m$ (Line 9 of Algorithm \ref{alg.6}) and 
based on Line 13 of Algorithm \ref{alg.3.p1}, $s_k$ has added $m$ to the list $DELIV\_MES_{k}$. Also $s_k$ has forwarded it to MH nodes in its group (Line 12 of Algorithm \ref{alg.6}), thus $m$ exists in the list $DELIV\_MES_{k}$. According to Line 21 of Algorithm \ref{alg.5}, if $h_i$ has not previously delivered the messages which exists in $DELIV\_MES_{k}$, $s_k$ forwards these messages to $h_i$. When $h_i$ handed off from $s_j$ to $s_k$, according to Line 5 of Algorithm \ref{alg.5}, $s_j$ sent the information belonging to $h_i$ to $s_k$. As a result, $s_k$ knows that $h_i$ has previously bcm-Hdelivered $m$, then $s_k$ does not forward $m$ to $h_i$. Thus, $h_i$ bcm-Hdelivers $m$ at most once from $s_j$.  \qed  

\begin{Lma} \label{lm.6}
    \normalfont If a non-faulty MH $h_i$ bcm-Hbroadcasts a message $m$, it bcm-Hdelivers $m$.
\end{Lma}
\noindent \textbf{Proof.} When $h_i$ bcm-Hbroadcasts an arbitrary message $m$, according to Line 6 of Algorithm \ref{alg.7}, it invokes the operation br-broadcast() to broadcast $m$ in the causal delivery management layer. Also, when $h_i$ intends to br-broadcast $m$, based on Line 1 of Algorithm \ref{alg.1}, it sends the message $m$ as \textit{INIT()} to all MH nodes in $VIEW_i$ and, the MSS node, $TELEPONT_i$. Since $h_i$ may be a member of the current group or may leave this group, we prove this lemma in two parts, which include: \textit{Part 1)} If $h_i$ is still in the current group, then according to Line 6 of Algorithm \ref{alg.1}, after receiving \textit{INIT()}, $h_i$ sends a message \textit{ECHO()} to $TELEPONT_i$, and it receives $m$ as \textit{READY()} from $TELEPONT_i$ (the second "When" of Algorithm \ref{alg.7}). Thus, $ h_i$ bcm-Hdelivers $m$ (Line 7 of Algorithm \ref{alg.7}). \textit{Part 2)} Consider that after bcm-Hbroadcasting $m$, $h_i$ disconnects from $s_j$ and joins $s_k$. Thus, when $s_j$ receives a message \textit{ECHO()} from $\frac{2NMH_k}{3}+1$ different MH nodes in its group (Line 6 of Algorithm \ref{alg.2}), it br-delivers $m$ in the Byzantine failure management layer (Line 10 of Algorithm \ref{alg.2}), then according to Line 2 of Algorithm \ref{alg.3.p1}, $s_j$ c-delivers $m$ in the causal delivery management layer. Based on Line 2 of Algorithm \ref{alg.6}, $s_j$ bcm-Sdelivers $m$ and invokes the operation c-broadcast() (Line 8 of Algorithm \ref{alg.6}). In Line 7 of Algorithm \ref{alg.3.p1}, $s_j$ sends $m$ to other MSS nodes (including $s_k$). According to Line 11 of Algorithm \ref{alg.3.p1} and Line 9 of Algorithm \ref{alg.6}, $s_k$ c-delivers and bcm-Sdelivers $m$, respectively. Finally, $s_k$ forwards $m$ to all MH nodes in its group (Line 12 of Algorithm \ref{alg.6}), and $h_i$ bcm-Hdelivers $m$ in Line 13 of Algorithm \ref{alg.7}. \qed
\begin{Theo}\label{theo.5.6}
    \normalfont The BCM-Broadcast algorithm satisfies the property BCM-Termination 1.
\end{Theo}
\noindent \textbf{Proof.} This theorem follows directly from Lemma \ref{lm.5} and Lemma \ref{lm.6}. \qed
\begin{Theo}\label{theo.5.7}
    \normalfont The BCM-Broadcast algorithm satisfies the property BCM-Termination 2.
\end{Theo}
\noindent \textbf{Proof.} When $s_j$ bcm-Sdelivers an arbitrary message $m$ from $h_i$ (Line 2 of Algorithm \ref{alg.6}), then $s_j$ c-broadcasts $m$ according to Line 8 of Algorithm \ref{alg.6}. Based on Line 7 of Algorithm \ref{alg.3.p1}, $s_j$ sends $m$ to $S_{j}\_VIEW$, where all MSS nodes in the system are members of $S_{j}\_VIEW$. Thus, an MSS $s_k$ c-delivers $m$ from $s_j$ (Line 11 of Algorithm \ref{alg.3.p1}) and bcm-Sdelivers (Line 9 of Algorithm \ref{alg.6}). Since all MSS nodes execute the same code for message delivery, all MSS nodes bcm-Sdeliver $m$ from $h_i$ through from $s_j$. \qed
\begin{Lma}\label{lm.7}
    \normalfont If a non-faulty MH $h_x$ bcm-Hdelivers a message $m$ from an MH $h_i$, where $h_x$ and $h_i$ are in the same group, then all non-faulty MH nodes in this group bcm-Hdeliver the message $m$ from $h_i$.
\end{Lma}
\noindent \textbf{Proof.} We know that the MH nodes in the desired group deliver exchanged messages in the system through the MSS $s_j$ to which they are connected. When $s_j$ bcm-Sdelivers an arbitrary message $m$ from $h_i$, based on Line 5 of Algorithm \ref{alg.6}, it sends $m$ as a message \textit{READY()} to all MH nodes in its group. An MH $h_x$ bcm-Hdelivers $m$ from $s_j$ in Line 7 of Algorithm \ref{alg.7}. Since all MH nodes execute the same algorithm, all MH nodes connected to $s_j$ receive a message \textit{READY()} from $s_j$, then bcm-Hdeliver $m$. \qed
\begin{Lma}\label{lm.8}
    \normalfont If a non-faulty MH $h_x$ bcm-Hdelivers a message $m$ from an MH $h_i$, where $h_x$ and $h_i$ are not in the same group, then all non-faulty MH nodes in other groups bcm-Hdeliver the message $m$ from $h_i$.
\end{Lma}
\noindent \textbf{Proof.} Consider an arbitrary MH $h_i$ in a group belonging to an MSS $s_j$. Based on Theorem \ref{theo.5.7}, all participating MSS nodes in the system bcm-Sdeliver a message $m$ from $h_i$ through $s_j$. Suppose an MH $h_x$ is connected to an MSS $s_k$. According to Line 9 of Algorithm \ref{alg.6}, when $s_k$ bcm-Sdelivers the message $m$ from $s_j$, Line 12 of Algorithm \ref{alg.6} forwards it to all MH nodes in its group. Finally, $h_x$ bcm-Hdelivers $m$ (Line 13 of Algorithm \ref{alg.7}). Since all MH and MSS nodes execute the same algorithm, all MSS nodes forward $m$ to the MH nodes in their groups; thus, all MH nodes bcm-Hdeliver $m$.  \qed
\begin{Theo} \label{theo.5.8}
    \normalfont The BCM-Broadcast algorithm satisfies the property BCM-Termination 3.
\end{Theo}
\noindent \textbf{Proof.} This theorem follows directly from Lemma \ref{lm.7} and Lemma \ref{lm.8}. \qed

\begin{Theo} \label{theo.5.9}
    \normalfont The BCM-Broadcast algorithm satisfies the property BCM-Termination 4.
\end{Theo}
\noindent \textbf{Proof.} Consider an arbitrary MH $h_i$ leaves an MSS $s_j$ and join another MSS $s_k$. Before the handing off, $h_i$ bcm-Hbroadcasts an arbitrary message $m$. When $h_i$ hands off to $s_k$, according Line 5 of Algorithm \ref{alg.5}, $s_j$ sends information about $h_i$ to $s_k$ as a message \textit{removed()}. Based on Theorem \ref{theo.5.7}, $s_k$ bcm-Sdelivers $m$. Upon delivering $m$, in Line 13 of Algorithm \ref{alg.3.p1}, $s_k$ adds it to the list $DELIV\_MES_{k}$. Then, after joining $h_i$ to $s_k$, $s_k$ checks the information about $h_i$ and whether it has delivered all the messages in the list $DELIV\_MES_{k}$ (Line 19 of Algorithm \ref{alg.5}); thus, $s_k$ forwards the messages that exist in the list $DELIV\_MES_{k}$ and $h_i$ has not bcm-Hdelivered yet (Line 21 of Algorithm \ref{alg.5}). Thus, $h_i$ bcm-Hdelivers $m$ in Line 10 of Algorithm \ref{alg.7}.\qed
\begin{Theo} \label{theo.5.10}
    \normalfont The BCM-Broadcast algorithm satisfies the property BCM-Causality 1.
\end{Theo}
\noindent \textbf{Proof.} To prove this theorem, consider the following scenario: 1) an arbitrary MH $h_i$ connected to an MSS $s_j$, bcm-Hbroadcasts a message $m_1$ in a group belonging to $s_j$ (the first “When” of Algorithm \ref{alg.7}), 2) after bcm-Sdelivering $m_1$, in Line 2 of Algorithm \ref{alg.6}, $s_j$ sends $m_1$ as a \textit{READY()} to the MH nodes of its group (Line 5 of Algorithm \ref{alg.6}), 3) $s_j$ c-broadcasts $m_1$ to other MSS nodes (Line 8 of Algorithm \ref{alg.6}), 4) after bcm-Hdelivering $m_1$, $h_i$ leaves $s_j$ and joins another MSS $s_k$, 5) $h_i$ bcm-Hbroadcasts the message $m_2$ in its group, 6) $s_k$ receives $m_2$ before bcm-Sdelivering $m_1$, 7) according to Line 4 of Algorithm \ref{alg.3.p1}, $s_k$ adds $m_2$ to the list $S_{k}\_RECEIVEDfromH$ and waits to deliver $m_1$, 8) based on Theorem \ref{theo.5.7}, $s_k$ eventually bcm-Sdelivers $m_1$, 9) $s_k$ bcm-Sdelivers $m_2$ (Line 2 of Algorithm \ref{alg.6}) after c-delivering $m_2$ from list  $S_{k}\_RECEIVEDfromH$ in Line 26 of Algorithm \ref{alg.3.p2}, 10) after bcm-Sdelivering $m_2$, $s_k$ sends $m_2$ as a \textit{READY()} to the MH nodes of its group (Line 5 of  Algorithm \ref{alg.6}) and also c-broadcasts $m_2$ to other MSS nodes in Line 8 of Algorithm \ref{alg.6}, 11) finally, $s_k$ bcm-Sbroadcasts $m_2$, where $m_2$ carries the message $m_1$ in its $cb$. When an arbitrary MSS $s_m$ receives $m_2$ before bcm-Sdelivering $m_1$, it checks the message's $cb$ content in Line 9 of Algorithm \ref{alg.3.p1}. Thus, $s_m$ bcm-Sdelivers $m_1$ before bcm-Sdelivering $m_2$, also $s_m$ first forwards $m_1$ and then forwards $m_2$ to the MH nodes in its group. Figure \ref{fig6} illustrates the execution of this scenario. \qed
\begin{Theo} \label{theo.5.11}
    \normalfont The BCM-Broadcast algorithm satisfies the property BCM-Causality 2.
\end{Theo}
\noindent \textbf{Proof.} Based on Line 1 of Algorithm \ref{alg.6}, when $s_j$ intend to bcm-Sbroadcast a message $m_1$, it invokes the operation c-broadcast() from the causal delivery management layer. Also, $s_j$ sends $m_1$ to all MSS nodes in Line 7 of Algorithm \ref{alg.3.p1}. Based on Theorem \ref{theo.5.7}, an arbitrary MSS $s_k$ bcm-Sdelivers any message that $s_j$ has broadcast. When an arbitrary MH $h_i$ hands off to $s_k$, two states may occur, which include: 1) before joining $h_i$, $s_k$ forwarded the messages $m_1$ and $m_2$ to MH nodes in its group in causal order. 2) $s_k$ forwards $m_1$ and $m_2$ to MH nodes in its group before and after $h_i$ joins $s_k$, respectively.

\noindent \textit{\underline{State 1}}: When $h_i$ connects to $s_k$, it sends a message \textit{requestMsg()} to $s_k$ (Line 6 of Algorithm \ref{alg.4}), then, according to Line 21 of Algorithm \ref{alg.5}, $s_k$ forwards the message $m_1$ and then $m_2$ to $h_i$ in the same order as delivered. Since communication channels support FIFO, $h_i$ first bcm-Hdelivers $m_1$ and then bcm-Hdelivers $m_2$ in Line 10 of Algorithm \ref{alg.7}.

\noindent \textit{\underline{State 2}}: Based on State 1, after connecting to $s_k$, $h_i$ bcm-Hdelivers $m_1$ in Line 10 of Algorithm \ref{alg.7}. Thus, when $s_k$ bcm-Sdelivers $m_2$, it forwards $m_2$ to MH nodes in its group, and $h_i$ bcm-Hdelivers $m_2$ in Line 13 of Algorithm \ref{alg.7}. As a result, $h_i$ first bcm-Hdelivers $m_1$ and then bcm-Hdelivers $m_2$.     \qed
\begin{Lma}\label{lm.9}
    \normalfont If an MH node first bcm-Hdelivers or bcm-Hbroadcasts  a message $m_1$ and then bcm-Hbroadcasts the message $m_2$, no MSS node bcm-Sdelivers $m_2$ before $m_1$.
\end{Lma}
\noindent \textbf{Proof.} When an arbitrary MH $h_i$ bcm-Hdelivers or bcm-Hbroadcasts a message $m_1$ in a group belonging to $s_j$, $h_i$ may face two cases to bcm-Hbroadcast a message $m_2$, which include: 1) $h_i$ bcm-Hbroadcasts $m_2$ in the same group that bcm-Hdelivers or bcm-Hbroadcasts $m_1$. 2) $h_i$ leaves $s_j$ and joins $s_k$, then bcm-Hbroadcasts $m_2$.

\noindent \textit{\underline{Case 1}}: When $h_i$ br-broadcasts $m_2$ in Line 6 of algorithm \ref{alg.7} and then sends it as a message \textit{INIT()} to MH nodes and MSS $s_j$ in its group (Line 1 of algorithm \ref{alg.1}), as well as when $s_j$ receives \textit{ECHO()} from  $\frac{2NMH_k}{3}+1$ different MH nodes in its group (Line 6 of Algorithm \ref{alg.2}), $s_j$ br-delivers the message $m_2$ (Line 10 of Algorithm \ref{alg.2}). In Line 1 of Algorithm \ref{alg.3.p1}, $s_j$ checks the status of delivering previous messages; if $s_j$ previously bcm-Sdelivered $m_1$, then $s_j$ also c-delivers $m_2$. Otherwise, $s_j$ temporarily stores $m_2$ in the buffer $S_{j}\_RECEIVEDfromH$. According to Line 23 of Algorithm \ref{alg.3.p2}, $s_j$ checks messages of $S_{j}\_RECEIVEDfromH$; if $s_j$ has bcm-Sdelivered $m_1$, then c-delivers $m_2$  in Line 18 of Algorithm \ref{alg.3.p2}, and finally, in Line 2 of Algorithm \ref{alg.6}, it bcm-Sdelivers $m_2$. Thus, $s_j$ bcm-Sdelivers $m_2$ after bcm-Sdelivering $m_1$. Since communication between the MSS nodes is FIFO, $s_j$ c-broadcasts $m_1$ before $m_2$ to other MSS nodes (Line 8 of Algorithm \ref{alg.6}). As a result, all MSS nodes bcm-Sdeliver $m_1$ before $m_2$.

\noindent \textit{\underline{Case 2}}: This State follows directly from Theorem \ref{theo.5.6}. \qed
\begin{Lma} \label{lm.10}
    \normalfont If an MSS node first bcm-Sdelivers or bcm-Sbroadcasts a message $m_1$ and then bcm-Sbroadcasts the message $m_2$, no MSS node bcm-Sdelivers $m_2$ before $m_1$.
\end{Lma}
\noindent \textbf{Proof.} We need to show two cases to prove this lemma.

\noindent \textit{\underline{Case 1}}: $s_k$ bcm-Sbroadcasts a message $m_1$ and then bcm-Sbroadcasts the message $m_2$. Since there is a FIFO channel between $s_k$ and all other MSS nodes, when $s_k$ bcm-Sbroadcasts $m_1$ and then bcm-Sbroadcasts $m_2$ (the first “When” of Algorithm \ref{alg.6}), thus an MSS $s_m$ bcm-Sdelivers $m_1$ before $m_2$ (Line 9 of Algorithm \ref{alg.6}). Based on Theorem \ref{theo.5.7}, all MSS nodes bcm-Sdeliver an arbitrary message from $h_i$ through from $s_j$; therefore, we conclude that all MSS nodes also bcm-Sdeliver an arbitrary message from $s_j$. Thus, all MSS nodes bcm-Sdeliver $m_1$ before $m_2$.

\noindent \textit{\underline{Case 2}}: $s_k$ bcm-Sdelivers a message $m_1$ from $s_j$ and then bcm-Sbroadcasts the message $m_2$. According to Line 10 of Algorithm \ref{alg.3.p1}, $s_k$ adds the message $m_1$ to its $cb$, then c-delivers it (Line 11 of Algorithm \ref{alg.3.p1}), and finally bcm-Sdelivers the desired message (Line 9 of Algorithm \ref{alg.6}). Afterward, $s_k$ bcm-Sbroadcasts $m_2$ in first “When” of Algorithm \ref{alg.6}. Consider an arbitrary MSS $s_m$ receives the message $m_2$ in third “When” of Algorithm \ref{alg.3.p1}, however, $s_m$  has not yet bcm-Sdelivered $m_1$. $s_m$ checks the content of $cb$ of $m_2$ in Line 9 of Algorithm \ref{alg.3.p1}. $s_m$ is informed that it has not yet bcm-Sdelivered $m_1$, then it temporarily stores $m_2$ in the buffer $S_{m}\_RECEIVEDfromS$ (Line 37 of Algorithm \ref{alg.3.p2}). Finally, when $s_m$ bcm-Sdelivers $m_1$, then it deletes $m_2$ from $S_{m}\_RECEIVEDfromS$ (Line 15 of Algorithm \ref{alg.3.p1}) and c-delivers, and finally, it bcm-Sdelivers $m_2$ (Line 9 of algorithm \ref{alg.6}). As a result, $s_m$ bcm-Sdelivers $m_1$ before $m_2$. Since all MSS nodes execute the same algorithm, all MSS nodes bcm-Sdeliver $m_1$ before $m_2$.  	\qed
\begin{Lma} \label{lm.11}
    \normalfont If an MH node first bcm-Hdelivers or bcm-Hbroadcasts  a message $m_1$ and then bcm-Hbroadcasts the message $m_2$, no non-faulty MH node bcm-Hdelivers $m_2$ before $m_1$.
\end{Lma}
\noindent \textbf{Proof.} Based on Lemma \ref{lm.9}, if an MH node first bcm-Hdelivers or bcm-Hbroadcasts a message $m_1$ and then bcm-Hbroadcasts the message $m_2$, all MSS nodes bcm-Sdeliver $m_1$ before $m_2$. We know that an arbitrary MH $h_i$ bcm-Hdelivers the broadcast messages in the system through an MSS to which it is connected. Therefore, an MH $h_i$ bcm-Hdelivers m$_1$ before $m_2$. Since all MH nodes execute the same algorithm, all MH nodes bcm-Hdeliver $m_1$ before $m_2$. \qed
\begin{Lma} \label{lm.12}
    \normalfont If an MSS node first bcm-Sdelivers or bcm-Sbroadcasts a message $m_1$ and then bcm-Sbroadcasts the message $m_2$, no non-faulty MH node bcm-Hdelivers $m_2$ before $m_1$.
\end{Lma}
\noindent \textbf{Proof.} Based on Lemma \ref{lm.10}, If an MSS node first bcm-Sdelivers or bcm-Sbroadcasts a message $m_1$ and then bcm-Sbroadcasts the message $m_2$, all MSS nodes bcm-Sdeliver $m_1$ before $m_2$. We know that an arbitrary MH $h_i$ bcm-Hdelivers the broadcast messages in the system through an arbitrary MSS to which it is connected. Thus, an MH $h_i$ bcm-Hdelivers $m_1$ before $m_2$. Since all MH nodes execute the same algorithm, all MH nodes bcm-Hdeliver $m_1$ before $m_2$. \qed      
\begin{Theo} \label{theo.5.12}
    \normalfont The BCM-Broadcast algorithm satisfies the property BCM-Causality 3.
\end{Theo}
\noindent \textbf{Proof.} This theorem follows directly from Lemmas \ref{lm.9}, \ref{lm.10}, \ref{lm.11}, and \ref{lm.12}. \qed
\begin{Theo} \label{theo.5.13}
    \normalfont The BCM-Broadcast algorithm satisfies the property BCM-Safety.
\end{Theo}
\noindent \textbf{Proof.} Consider a Byzantine MH node in Line 1 of Algorithm \ref{alg.1} that sends an arbitrary message $m$ as a message \textit{INIT()} in its group, the message $m$ may be either 1) a duplicate message, or 2) a malicious message.

\noindent \textit{\underline{Case 1}}: When an arbitrary MSS $s_j$ as a \textit{TELEPOINT} of MH $h_i$, receives a message \textit{INIT()}, then checks whether the message is duplicated (Line 1 of Algorithm \ref{alg.2}). If $m$ is a duplicate message, $s_j$ ignores $m$ according to Line 4 of Algorithm \ref{alg.2}. Also, according to Line 3 of Algorithm \ref{alg.1}, a non-faulty MH node from $h_i$'s group checks the duplication of the message \textit{INIT()}. If $m$ is a duplicate message, the MH node disregards it (Line 4 of Algorithm \ref{alg.1}).

\noindent \textit{\underline{Case 2}}: According to Line 6 of Algorithm \ref{alg.1}, all MH nodes in an arbitrary group send a message \textit{ECHO()} to \textit{TELEPONT} (e.g., $s_j$) after receiving the message \textit{INIT()} from $h_i$. Ideally, $s_j$ receives \textit{ECHO()} from at least $\frac{2NMH_j}{3}+1$ different MH nodes in its group. Consider $s_j$ receives $ECHO(\left<i, seq_{i}\right>,m)$ from $h_y$ and $ECHO(\left<i, seq_{i}\right>,m^{'})$ from $h_x$, where the message received in \textit{ECHO()} belonging to MH $h_y$ may be different from the message received in \textit{ECHO()} belonging to MH $h_x$ (Line 6 of Algorithm \ref{alg.2}); Thus, $s_j$ disregards the message \textit{INIT()} and the received messages \textit{ECHO()} (Line 8 of Algorithm \ref{alg.2}). Since MH nodes in a group deliver messages through their \textit{TELEPOINT}, no non-faulty MH node bcm-Hdelivers malicious message; because $s_j$ ignores malicious message. \qed

\begin{Theo} \label{theo.5.90}
\normalfont If three messages, $m_1$, $m_2$, and $m_3$, have a causal relationship, then all MH nodes bcm-Hdeliver them in causal order; if an arbitrary MH $h_i$ does not bcm-Hdelivers one of the messages and violates the causal order, $h_i$ is a Byzantine node.
\end{Theo}
\noindent \textbf{Proof.} Consider the MH nodes $h_x$, $h_y$, and $h_i$ connected to an arbitrary MSS $s_j$ in an arbitrary group. $h_x$ bcm-Hbroadcasts the messages $m_1$ and then $m_2$. Consider $h_x$ first bcm-Hdelivers $m_1$ and then bcm-Hbroadcasts $m_2$. Thus, $m_1$ and $m_2$ have a cause-and-effect relationship. $h_y$ first bcm-Hdelivers $m_1$ and $m_2$, respectively; then bcm-Hbroadcasts $m_3$. Thus, the MSS $s_j$ (respectively, the non-faulty MH nodes) should bcm-Sdeliver (respectively, bcm-Hdeliver) $m_1$ and $m_2$ before $m_3$. We prove this by contradiction. Based on Theorem \ref{theo.5.8}, if a non-faulty MH $h_y$ has bcm-Hdelivered an arbitrary message, then all non-faulty MH nodes bcm-Hdelivers it. Thus, all non-faulty MH nodes bcm-Hdeliver $m_1$ and $m_2$ before $m_3$. Consider an arbitrary MH $h_i$ first bcm-Hdelivers $m_1$ and then does not bcm-Hdeliver $m_2$ before bcm-Hdelivering $m_3$. According to Line 5 of Algorithm \ref{alg.6}, $s_j$ send the message \textit{READY($m_2$)} to all MH nodes in its group. According to Line 7 of Algorithm \ref{alg.7}, $h_i$ bcm-Hdelivers $m_2$ from $s_j$. Even if $h_i$ has just joined the group, according to Line 21 of Algorithm \ref{alg.5}, $s_j$ forwards $m_2$ to $h_i$. Since $s_j$ forwards the messages in the same order as bcm-Sdelivered to MH nodes in its group, it has forwarded $m_2$ before forwarding $m_3$. Thus, a non-faulty MH $h_i$ bcm-Hdelivers $m_2$ before $m_3$. If $h_i$ still has not bcm-Hdelivered $m_2$ before bcm-Hdelivering $m_3$, then $h_i$ is a Byzantine node, and our algorithm does not guarantee the preservation of causal order for Byzantine nodes.

\textit{Analysis of the message complexity}: The BCM-Broadcast algorithm is optimal concerning t-resilience (i.e., $t<\frac{N_{mh}}{3}$). Consider that an arbitrary MH $h_i$ is connected to an MSS $s_j$; our algorithm requires three consecutive communication steps and $3NMH_{j}+1$ messages to exchange messages between $h_i$ and MH nodes in its group, where $NMH_{j}+1$ messages \textit{INIT()}, $NMH_j$ messages \textit{ECHO()}, and $NMH_j$ messages $READY()$, which communication steps include: 

\noindent \textit{Step 1)} $h_i$ sends $NMH_{j}+1$ messages $m$ as a message \textit{INIT()} to all MH nodes in its group and the MSS $s_j$.

\noindent \textit{Step 2)} all MH nodes in the group send $NMH_j$ messages \textit{ECHO()} to $s_j$.

\noindent \textit{Step 3)} $s_j$ sends $NMH_j$ messages \textit{READY()} to all MH nodes in its group. 

Also, BCM-Broadcast algorithm requires four consecutive communication steps and $(2NMH_{j}+1)+N_{mss}+(N_{mh}-NMH_{j}) = NMH_{j}+N_{mss}+N_{mh}+1$ messages to exchange messages between $h_i$ and MH nodes in other groups, which communication steps include:

\noindent \textit{Step 1)} $h_i$ sends an arbitrary message $m$ as a message \textit{INIT()} to all MH nodes in its group and $s_j$.

\noindent \textit{Step 2)} all MH nodes in the group send a message \textit{ECHO()} to $s_j$.

\noindent \textit{Step 3)} $s_j$ bcm-Sbroadcast $m$ to all MSS nodes in systems.

\noindent \textit{Step 4)} all MSS nodes forward $m$ to MH nodes in its group. 

\noindent The message complexity of our algorithm is $O(n)$, compared to Bracha \cite{3} and Imbs-Raynal \cite{4} algorithms, both of which are $O(n^{2})$.
\section{Stochastic Analysis of the t-Condition}\label{sec:analysis}
This section first introduces each behavior of an MH node as an event. Then, it analyzes the probability of the violation of the t-condition based on the probability of each event. Due to the Byzantine and mobile nature of MH nodes, if a Byzantine MH node joins an arbitrary group or a non-faulty MH node leaves the group, in both cases, the total number of Byzantine processes in the desired group may exceed one-third of the total number of MH nodes; thus, these events may violate the t-condition. Therefore, the correctness of the BCM-Broadcast algorithm depends on meeting the condition that the number of Byzantine nodes is less than a third of the total number of mobile nodes. In the rest of this section, we analyze the probability of the violation of the safety condition under different scenarios.

\textit{Occurrence of different events in a group}: Let us assume that $e_x$ is an event that represents the behavior of a (non-faulty/Byzantine) node in the group, where $x = \left\{1,2,3,4\right\}$. The events include:
\begin{itemize}
    \item $e_1$ represents that a Byzantine MH node leaves an arbitrary group.
    \item $e_2$ represents that a non-faulty MH node leaves an arbitrary group.
    \item $e_3$ represents that a Byzantine MH node joins an arbitrary group.
    \item $e_4$ represents that a non-faulty MH node joins an arbitrary group.
\end{itemize}
Consider an arbitrary group where any of these events may occur simultaneously. We assume that $k_1$ to $k_4$ are the number of occurrences of each event $e_1$ to $e_4$, respectively. Also, we assume that the total number of events in the group equals $E$. For example, when $4$ Byzantine nodes join the group, $k_{3} = 4$ and $5$ non-faulty nodes leave the group, $k_{2} = 5$. Thus, we have $E = k_{2} + k_{3} = 9$.

\textit{Calculate the average rate of each event}: Consider one or more events mentioned above that occur in the time interval $[t,t+\tau)$, where $\tau$ represents the specified period of time. The parameter $\lambda_x$ indicates the average rate of event $e_x$ in $\tau$, and $k_x$ is the number of occurrences of event $e_x$ in $\tau$, which  $\lambda_x$ is obtained from (\ref{eq.1}). For example, if $2$ Byzantine nodes leave the arbitrary group in $24$ minutes, then the average rate of leaving a group by Byzantine nodes is $0.08$ $(\lambda_{x}=\frac{2}{24}=0.08)$.
\begin{equation}\label{eq.1}
     \lambda_{x} = \frac{k_{x}}{\tau}
\end{equation}
Thus, $\lambda_1$ to $\lambda_4$ represent the rate of occurrences of $e_1$ to $e_4$ in $\tau$, respectively. To analyze the probability of the violation of the t-condition based on the probability of an event in a random time interval, we need to calculate the average rate of this event in the desired interval of time. For example, if $2$ Byzantine nodes leave an arbitrary group in $24$ minutes, according to (\ref{eq.2}) on average, $5$ Byzantine nodes leave the group every hour; thus, we have:  $\lambda_{1}=\frac{2}{24}\times{60} = 5$, where $\tau= 24$ and $\tau_{1}=60$.
\begin{equation}\label{eq.2}
     \lambda_{x} = \frac{k_{x}}{\tau}\times{\tau_{1}}
\end{equation}

\textit{The probability of the violation of the t-condition based on the probability of the occurrence of an event in a group}: Consider only one of the events $e_1$ to $e_5$ occurs in a period of time. If one of the events $e_1$, $e_4$, or $e_5$ occurs, the probability of the violation of the t-condition based on the probability of increasing Byzantine nodes in the group is zero. In contrast, if one of the events $e_2$ or $e_3$ occurs, then the probability of the violation of the t-condition based on the probability of increasing Byzantine nodes in the group is greater than zero. Scenarios \ref{sce.6.1} and \ref{sce.6.2} analyze the probability of occurrences of events $e_2$ and $e_3$, respectively.

\newtheorem{Sce}{Scenario}[section]
\begin{Sce}\label{sce.6.1}
    \normalfont Consider in a period of time $[t,t+\tau)$, $k_2$ non-faulty nodes leave an arbitrary group with rate $\lambda_2$. According to the Poisson distribution, $P(E=k_{2})$ in (\ref{eq.3}) represents the probability of $k_2$ non-faulty nodes leaving a group in the period of time $\tau$ with rate $\lambda_2$, and the act of leaving causes the violation of the t-condition:
\end{Sce}
\begin{equation}\label{eq.3}
    P(E = k_{2}) = \frac{e^{-(\tau \lambda_{2})}.{(\tau \lambda_{2})^{k_{2}}}}{k_{2}!}
\end{equation}
Thus, if $k_{2}  \geq NMH_{j} - 3t_{j}$, the probability of the violation of the t-condition is equal to $1$; if $k_{2} < NMH_{j} - 3t_{j}$, the probability of the violation of the condition is smaller than $1$, this means that the t-condition still holds; where, $NMH_j$ and $t_j$ respectively denote the number of MH nodes connected to an arbitrary MSS $s_j$ in the desired group and the number of Byzantine processes connected to the group, respectively.
\begin{Sce}\label{sce.6.2}
    \normalfont Consider in a period of time $[t,t+\tau)$, $k_3$ Byzantine nodes join an arbitrary group with rate $\lambda_{3}$. According to the Poisson distribution, $P(E = k_{3})$ in (\ref{eq.4}) represents the probability of $k_3$ non-faulty nodes joining the desired group in the period of time $\tau$ with rate $\lambda_{3}$, and the act of joining causes the violation of the t-condition:
    \begin{equation}\label{eq.4}
        P(E = k_{3}) = \frac{e^{-(\tau \lambda_{3})}.{(\tau \lambda_{3})^{k_{3}}}}{k_{3}!}
    \end{equation}
\end{Sce}
Thus, if $k_{3}  \geq \lfloor \frac{NMH_{j}}{3} \rfloor - t_{j}$, the probability of the violation of the t-condition is equal to $1$; if $k_{3}  < \lfloor \frac{NMH_{j}}{3} \rfloor - t_{j}$, the probability of the violation of the condition is smaller than $1$, this means that the t-condition still holds.

\textit{The occurrence of two or more simultaneous events}: The events $e_1$ to $e_4$ are independent, and two, three, or four simultaneous events may occur in a period of time $\tau$. According to (\ref{eq.5}), the total number of combination of events in the period of time $\tau$ is equal to $15$.
\begin{equation}
    \begin{split}
       \label{eq.5}
    \textit{Total number of combination of events} = \left ( \begin{array}{c} 4 \\ 1 \end{array} \right ) + \left ( \begin{array}{c} 4 \\ 2 \end{array} \right ) \\
    + \left ( \begin{array}{c} 4 \\ 3 \end{array} \right ) + \left ( \begin{array}{c} 4 \\ 4 \end{array} \right ) = 15   
    \end{split}
\end{equation}
    
Thus, if a series of events occur with different rates, we use (\ref{eq.6}) to calculate the probability of the violation of the t-condition based on the probability of the occurrence of different events in a group, where $E = k_{1} + k_{2} + k_{3} + k_{4}$ and $\lambda = \lambda_{1} + \lambda_{2} + \lambda_{3} + \lambda_{4}$.
\begin{equation}\label{eq.6}
    P(E) = \prod_{i=1}^4 \frac{e^{-(\tau\lambda_{i})}.(\tau\lambda_{i})^{k_{i}}}{k_{i}!}=\frac{e^{-\left(4\tau\lambda \right)}.\left(\tau^{4}\left(\lambda_{1}^{k_{1}}\lambda_{2}^{k_{2}}\lambda_{3}^{k_{3}}\lambda_{4}^{k_{4}}\right)\right)}{k_{1}!k_{2}!k_{3}!k_{4}!}
\end{equation}
\begin{Sce}\label{sce.6.3}
\normalfont Consider in a period of time $[t,t+\tau)$, $k_2$ non-faulty nodes leave an arbitrary group with rate $\lambda_2$ , where $k_{2}< \frac{2NMH_{j}}{3}$. Also, $k_1$ Byzantine nodes leave the desired group with rate $\lambda_1$, where $k_{1}\leq t_j$; and also, $k_3$ Byzantine nodes join the group with rate $\lambda_3$, where $k_{3} \geq \lfloor\frac{NMH_{j}-k_{1}-k_{2}}{3}\rfloor-t_{j}-k_{1}$ and $K_3$ is more than the number of non-faulty MH nodes remaining in the group. According to (\ref{eq.6}), $P(E)$ is the probability of violation of the t-condition based on the probability of the occurrence of the events $e_1$, $e_2$, and $e_3$, where $E = k_{1}+k_{2}+k_{3}$ and $\lambda = \lambda_{1}+\lambda_{2}+\lambda_{3}$.
\end{Sce}
\textit{The effect of the probability of violation of the t-condition on the probability of message loss in the group}: Scenario \ref{sce.6.2} violates the t-condition with probability $P(E=k_{3})$. Consider an arbitrary $h_i$ connected to the MSS $s_j$. Suppose $h_i$ broadcasts a message $m$ as \textit{INIT()} in the desired group of Scenario \ref{sce.6.2}. Since the total number of Byzantine processes in this group exceeds one-third of the total number of MH nodes, $s_j$ does not receive \textit{ECHO()} from  $\frac{2NMH_{j}}{3}+1$ different MH nodes in its group. Thus, $s_j$ refuses to confirm $m$ and sends \textit{READY()} to MH nodes in its group. Thus, in this scenario, no MH node bcm-Hdelivers $m$, and eventually, $m$ is lost. We know that $S_{j}\_DELIV[i]$ is the number of messages that $s_j$ has bcm-Sdelivered from $h_i$, and $seq_i$ represents the number of messages that $h_i$ bcm-Hbroadcast until now. $M$ represents the number of messages lost during this scenario in a period of time $\tau$, where $M = seq_{i}-S_{j}\_DELIV[i]$. The rate of sending messages by $h_i$ in the group at time $\tau$ equals $\epsilon$. According to the Poisson distribution in (\ref{eq.7}), $P(M)$ is the probability of losing some messages in a period of time $\tau$ based on Scenario \ref{sce.6.2}.
\begin{equation}\label{eq.7}
    P(M) = \frac{e^{-(\tau\epsilon)}.(\tau\epsilon)^{M}}{M!}.P(E=k_{3})
\end{equation}

\textit{The effect of the probability of message loss on the guarantees of causal order in the group}: The BCM-Broadcast algorithm's main goal is to guarantee messages' causal order. According to the assumption of the algorithm, the communication channels between MH nodes and MSS nodes are FIFO. Consider an MH $h_i$ bcm-Hbroadcasts the message $m_1$ and then $m_2$ or bcm-Hdelivers $m_1$ and then bcm-Hbroadcasts $m_2$, $h_i$ does not need to bcm-Hbroadcast the information of the previously delivered messages along with the messages. Since the MH and MSS nodes use FIFO channels, the MSS $s_j$ first bcm-Sdelivers $m_1$ and then $m_2$; thus, if the t-condition holds in the group and $s_j$ receives \textit{ECHO()} from  $\frac{2NMH_j}{3}+1$ different MH nodes, first it bcm-Sbroadcasts \textit{READY($m_1$)} and then \textit{READY($m_2$)}. As a result, all MH nodes bcm-Hdeliver $m_1$ before $m_2$.  Theorem \ref{theo.6.1} proves that the BCM-Broadcast algorithm guarantees the causal order in the other messages, even if a message is lost based on the violation of the t-condition.

\begin{Theo}\label{theo.6.1}
    \normalfont The loss of an arbitrary message does not affect the causal delivery of other messages. This means the BCM-Broadcast algorithm still guarantees the causal order of other messages even when the t-condition is violated and several messages are lost.
\end{Theo}
\noindent \textbf{Proof.} Consider the MH nodes $h_x$, $h_y$, and $h_i$ connected to an arbitrary MSS $s_j$ in an arbitrary group. $h_x$ bcm-Hbroadcasts the messages $m_1$ and then $m_2$. Also, $h_y$ bcm-Hbroadcasts the message $m_3$. Consider $h_x$ first bcm-Hdelivers $m_1$ and then bcm-Hbroadcasts $m_2$. Thus, $m_1$ and $m_2$ have a cause-and-effect relationship. Consider that the t-condition is violated in the group and a non-faulty MH $h_y$ loses some of the exchanged messages; however, it still preserves the causal order. Let us prove this by contradiction and assume that violating the t-condition causes a violation of the causal delivery between MH nodes. Consider $h_y$ bcm-Hdelivers $m_1$ but does not bcm-Hdelivers $m_2$, and $h_y$ bcm-Hbroadcasts $m_3$. Consider that the t-condition is violated in the group; therefore, $s_j$ does not receive \textit{ECHO($m_2$)} from  $\frac{2NMH_{j}}{3}+1$ different MH nodes in its group. Thus, $s_j$ does not send \textit{READY($m_2$)} and according to Line 13 of Algorithm \ref{alg.2}, $s_j$ disregards \textit{INIT($m_2$)}; thereby, no MSS  node (respectively, no non-faulty MH nodes) bcm-Sdelivers (respectively, bcm-Hdeliver) the message $m_2$, this means that $m_2$ is lost. Thus, we conclude that $h_y$ has not delivered $m_2$ before bcm-Hbroadcasting $m_3$, and there is no causal relationship between $m_2$ and $m_3$. Thus, there is no requirement that $s_j$ (respectively, the non-faulty MH nodes) bcm-Sdelivers (respectively, bcm-Hdeliver) $m_2$ before $m_3$, and to guarantee the causal order between messages, it is sufficient that $s_j$ (respectively, the non-faulty MH nodes) bcm-Sdelivers (respectively, bcm-Hdeliver) $m_1$ before $m_3$. \qed

\begin{figure}
    \centering
    \includegraphics[width=1\linewidth]{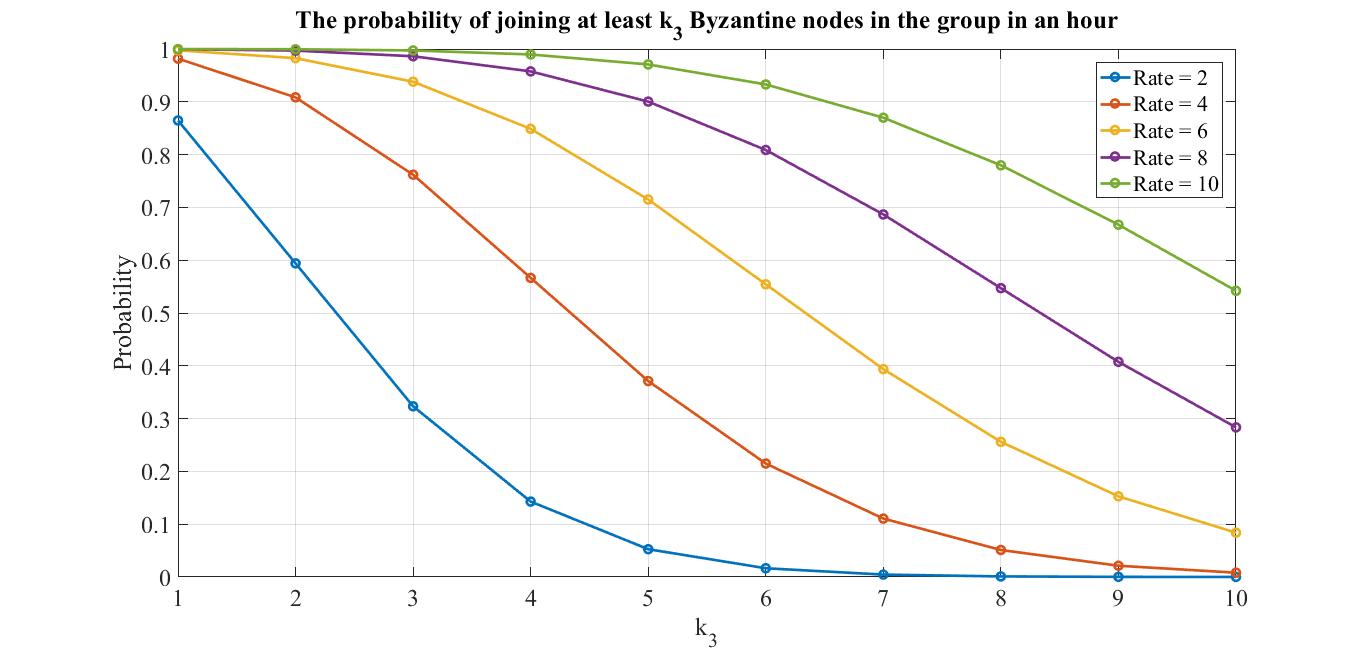}
    \caption{The probability of joining at least $k_3$ Byzantine nodes in the group in an hour with different rate values}
    \label{fig7}
\end{figure}

\textit{Case analysis of a scenario}: Consider an arbitrary group consisting of an MSS $s_j$ and several MH nodes. In a special period of time, the number of participating MH nodes is $NMH_{j}=30$. Ideally, the maximum number of Byzantine nodes should equal $t_{j}<10$. Consider that at this moment $t_{j}=7$; also, the Byzantine nodes join the group at a rate of $2$ nodes per $15$ minutes; therefore, Eq. (\ref{eq.8}) calculates the probability of violation of the t-condition based on the probability of joining at least $3$ Byzantine nodes in the group in an hour. We calculate the Byzantine node's joining rate to the group from (\ref{eq.2}), and we have:  $(\lambda_{3}=\frac{2}{15}\times 60 = 8)$.
\begin{equation}\label{eq.8}
    P(k_{3}\geq3)=1-P(k_{3}<3)=1-{\sum_{i=0}^{2}{\frac{e^{(-8)}.8^{i}}{i!}}}=0.9862
\end{equation}
With a probability close to $1$, at least $3$ Byzantine nodes join the group within an hour. Let's examine this scenario for different values of the joining rate. Figure \ref{fig7} illustrates the probability of joining at least $k_3$ Byzantine nodes in the group in an hour with different rate values. We conclude that increasing the rate value increases the probability of joining Byzantine nodes. 
Keeping the above assumptions, we assume that $h_i$ bcm-Hbroadcast 5 messages in an arbitrary group in half an hour. The number of messages that $s_j$ bcm-Sdelivered from $h_i$ in an hour is $14$. Equation (\ref{eq.9}) calculates the probability of losing at least 6 messages in an hour due to a violation of the t-condition. We calculate the message loss rate in a group from (\ref{eq.2}), and we have $\epsilon = \frac{5}{30}\times60=10$.
\begin{equation}\label{eq.9}
    P(M\geq 6) = 1-{P(M<6).P(k_{3}\geq3)=(1-\sum_{i=0}^{5}{\frac{e^{-10}.10^{i}}{i!})}}.P(k_{3}\geq3)
\end{equation}
We calculate the message loss probability for different values of $k_3$, assuming $(\lambda_{3}=8,\epsilon=10)$. Table \ref{tbl.2} illustrates that the probability of joining at least $k_3$ Byzantine MH node(s) to an arbitrary group directly correlates with the probability of message loss. As the probability of joining at least $k_3$ Byzantine MH node(s) to an arbitrary group increases, the probability of losing $M$ messages also increases and vice versa.
\def\arraystretch{1.0}
\begin{table}
    \caption{Correlation between the probability of joining at least $k_3$ Byzantine nodes and the probability of message loss (where $\lambda_{3}=8,\epsilon=10$)}\label{tbl.2}
    \centering
    \begin{tabular}{ccccc}
    \hline
$k_{3}$ & P($E\geq k_{3}$)& P($M=12$)&P($M=14$)&P($M=16$)\\
 \hline
1   &  0.999665 & 0.094749 & 0.052060 & 0.021692 \\ 
   2   &  0.996981 & 0.094494  & 0.051920 & 0.021633 \\ 
   3   &  0.986246 & 0.093477  & 0.051361 & 0.021400 \\
   4   &  0.957620 & 0.090764  & 0.049870 & 0.020779 \\ 
   5   &  0.900368 & 0.085337  & 0.046889 & 0.019537 \\ 
   6   &  0.808764 & 0.076655  & 0.042118 & 0.017549 \\ 
   7   &  0.686626 & 0.065079  & 0.035757 & 0.014899 \\ 
   8   &  0.547039 & 0.051849  & 0.028488 & 0.011870 \\ 
   9   &  0.407453 & 0.038618  & 0.021219 & 0.008841 \\ 
  10   &  0.283376 & 0.026858  & 0.014757 & 0.006149 \\ 
\hline
    \end{tabular}
\end{table}
\section{Related Works and Discussion}\label{sec:related}
In this section, we first review the history of causal broadcast algorithms and then introduce the works that have presented the causal broadcast algorithm for distributed mobile systems. In addition, we introduce works that have discussed the guarantee of causal order in the presence of Byzantine processes. Finally, we state how our algorithm can be inspired by previous algorithms.

\textit{The background of causal broadcast algorithms}: Birman \textit{et al.} introduced the causal broadcast algorithm for the first time in 1987 \cite{5} and used it in the ISIS project \cite{6}. Effective techniques have been introduced to guarantee the causal delivery of messages in distributed systems. The most famous of these techniques include: 1) The local broadcast message carries the delivered messages in the attachment \cite{7}, 2) The local broadcast message carries the meta-information of delivered messages in the attachment \cite{7, 34}, 3) Messages are tagged with a vector clock; this tagging informs the receiver process that there is another message that has not yet been delivered by receiver process \cite{8}. Techniques one and two are known as \textit{sequence numbers}, and technique three is known as \textit{vector clock}.

Since 1987, researchers developed many causal broadcast algorithms to use in different applications, inspired by the first proposed algorithm. These applications include: communication between processes of distributed systems \cite{2,14, 9,10,32,33,12,13}, guaranteeing causal consistency in distributed databases, read-write memory \cite{15, 16, 17}, and message exchange between nodes in mobile networks \cite{19,21,18,20,22,23,35}. Most past algorithms have been presented in the presence of non-faulty processes. Recently, some works have been presented that guarantee the causal order in the presence of Byzantine processes \cite{2, 14, 24, 25, 26, 27}. In the following, we will discuss these researches.

\textit{Previous Works Related to the Causal Order Algorithms for Distributed Mobile Systems}: Prakash \textit{et al.} \cite{28} have presented two efficient approaches to using vector clocks to maintain causal dependencies between processes in distributed mobile systems. The first approach is called \textit{dependency sequence}, and the second is the \textit{hierarchical clock}. Their desired system's architecture is the same one presented in \cite{1}. They call a group of several MH nodes with one MSS a \textit{cell}. In the first approach, an MSS node in an arbitrary cell stores the events that occur in that cell by the MH nodes in a vector. The number of components of the vector equals the number of MH nodes. For every event that occurs by an MH node, the component's value corresponding to that MH node in the vector increases. When an MSS node intends to broadcast an arbitrary message to other MSS nodes, it attaches this vector to the broadcast message. In the second approach, they have defined a hierarchical clock called $\varphi$, which includes $\phi^i$ and $\phi^m$. The first denotes the \textit{local clock}, and the second represents the \textit{global clock}. In the second approach, an MSS node attaches the hierarchical clock to the broadcast message instead of the vector. The first approach has a higher communication cost than the second approach, and the time required to build the history of dependencies in the first approach is less than in the second approach. Kshemkalyani \textit{et al.} \cite{12} have presented an algorithm called \textit{KS (Kshemkalyani and Signal)}, which guarantees causal order multicast for stationary systems, and this algorithm is optimal in terms of communication overhead. Chandra \textit{et al.} \cite{22} have developed this algorithm for cellular networks with mobile nodes, and they have compared their algorithm with the algorithm calles \textit{RST (Raynal, Schiper, and Toueg)} \cite{11} and also proved that their algorithm has less communication overhead than the RST algorithm. 

Benzaid \textit{et al.} \cite{29} have presented a protocol called \textit{Mobi\_Causal} to implement causal ordering in mobile computing systems, using two approaches proposed in \cite{28}. They have proved that the protocol they presented has low communication overhead and high scalability. Also, they have proved that their protocol satisfies the safety (this means that the causal order is never violated) and liveness (it means that each message is eventually delivered after a limited time). Their protocol is suitable for unicast communication. In separate research, Benzaid \textit{et al.} \cite{30} have presented another protocol called \textit{BMobi\_Causal} inspired by \cite{29} to guarantee the causal order of messages in broadcast communication. They have compared their protocol with other existing causal broadcast communication protocols and proved that the BMobi\_Causal is more optimal than others in terms of message overhead, storage overhead, and communication complexity. In addition, they have proved that the BMobi\_Causal has no delay in delivering messages due to guaranteeing the causality condition.

N´edelec \textit{et al.} \cite{19} have focused on the scalability of causal broadcast algorithms in large and dynamic systems. They have selected the algorithm and architecture of the mobile system used in \cite{31} as the basic algorithm. First, by adding new nodes to the mobile system \cite{31}, they have proven that increasing the number of nodes violates the causal order of messages. Then, by developing the algorithm presented in \cite{31}, they have provided a non-blocking causal broadcast protocol suitable for large and dynamic systems and efficient in terms of execution time complexity, local storage space complexity, and message size. Guidec \textit{et al.} \cite{21} have presented two algorithms based on the causal barriers that can guarantee the causal delivery of messages in opportunistic networks. They have considered the broadcast messages without a lifetime in the first algorithm, in contrast, every message has a lifetime in the second algorithm. As such, after the lifetime of a message, the information related to it and all the existing dependencies are deleted from the network. They have not provided any approach to manage the mobility of nodes.

 Most of the reviewed algorithms have used the vector clock technique to ensure the causal order of messages. These vector clocks have a fixed size. Wilhelm \textit{et al.} \cite{20} have introduced a dynamic clock set, which consists of a group of possible clocks that change the size of the elements of these vector clocks as the number of nodes in the system changes. While they have not introduced a special algorithm to guarantee the causal ordering of messages or a special technique to manage the mobility of nodes, their approach is efficient for dynamic systems where the number of nodes is constantly changing. In the reviewed works, some of them have only developed a new approach without presenting an algorithm \cite{20, 28}, and others have considered their algorithm in the architecture of mobile systems introduced in \cite{1} with the non-faulty processes \cite{30, 1, 29}. Recently, researchers have presented other works on guaranteeing the causal order of messages in the presence of Byzantine processes, which processes are stationary in their target system, which we will discuss below.
 
\textit{Previous Works Related to the Causal Order Algorithms in the Presence of Byzantine Processes:} Until 2021, the researchers have presented no causal broadcast algorithm in the presence of Byzantine processes. Auvate \textit{et al.} \cite{2} have represented a modular \textit{BCO-Broadcast} algorithm for this aim for the first time. Thus, the BCO-Broadcast algorithm is implemented on top of the BR-Broadcast algorithm \cite{3}, and the proof of the properties of \textit{BCO-Validity}, \textit{BCO-Integrity}, \textit{BCO-Termination 1}, \textit{BCO-Termination 2}, and \textit{BCO-Causality} are inspired by BR-Broadcast properties. Misra \textit{et al.} \cite{14} have chosen the presented protocol in \cite{11} and, by defining an attack, prove that the algorithm fails under Byzantine attacks and shows that exchanged messages under point-to-point communications in the asynchronous system still retain causal order. Misra \textit{et al.} \cite{25} have investigated one of the most popular protocols for a reliable broadcast under the Byzantine failure presented by Bracha \cite{3} and also the BR-Broadcast algorithm presented by Imbs-Raynal \cite{4}. Also, they have examined the guarantee or non-guarantee of the causal order of these two protocols for unicast and multicast communications in the absence and presence of Byzantine processes \cite{24, 26}. Since it is possible to have Byzantine processes in any mobile system, but so far, no algorithm has been presented that can guarantee the causal order in the presence of Byzantine and mobile processes.
\section{Conclusions and Future Work}\label{sec:concl}

We presented a layered BCM-Broadcast algorithm for the implementation of causal broadcast in distributed mobile networks. BCM-Broadcast has a modular and hierarchical architecture, where there is a fixed number of  MH and  MSS nodes. The MSS nodes are stationary and non-faulty. By contrast,  the MH nodes could move and may become Byzantine. The BCM-Broadcast algorithm, using three layers of the Byzantine failure management layer, the causal delivery management layer, and the mobility management layer, simultaneously manages the three critical challenges of Byzantine failure, causal delivery, and mobility. First, we defined twelve  important properties for the BCM-Broadcast algorithm and then proved that it satisfies all these properties, including its safety.  Moreover, using the Poisson process, we analyzed the probability of violating the t-condition (i.e., the number of Byzantine nodes is less than a third of the total number of nodes) and the probability of losing  messages under different mobility scenarios. Further, we showed that the message complexity of our algorithm is linear in the number of nodes. (Most existing methods such as  \cite{3} and  \cite{4} have a quadratic message complexity.) 

We plan to extend this work in at least the following three directions. First, we will investigate a completely mobile architecture, where all nodes can move. Second, we will investigate a scenario, where both MH and MSS nodes can be Byzantine. Third, we will extend the BCM-Broadcast algorithm for a system whose number of nodes constantly changes as the number of nodes decreases/increases due to leaving/joining in the system.


\end{document}